\documentclass[conference]{IEEETran}
\usepackage[utf8]{inputenc}
\usepackage{graphicx}
\usepackage{amsmath}
\usepackage{amsfonts}
\usepackage{booktabs}
\usepackage{multirow}
\usepackage{wrapfig}
\usepackage[hidelinks]{hyperref}
\usepackage{dblfloatfix}
\usepackage{cleveref}
\usepackage[numbers,sort&compress]{natbib}
\usepackage{amsthm}
\usepackage{subcaption}
\usepackage{bbm}
\usepackage{algorithmic}
\usepackage{algorithm}
\usepackage{newfloat}
\usepackage{wasysym}
\usepackage{xcolor}

\newcommand{\algcomment}[1]{\hfill {\color{blue} $\triangleright$ \emph{\small{#1}}}}

\title{Membership Inference Attacks From First Principles}

\author{Nicholas Carlini$^{*1}$  \quad Steve Chien$^1$  \quad Milad Nasr$^{1,2}$  \quad Shuang Song$^1$  \quad Andreas Terzis$^1$  \quad Florian Tramèr$^1$  \\ \emph{$^1$ Google Research \quad $^2$ University of Massachusetts Amherst}}

\begin{document}

\maketitle

\thispagestyle{plain}
\pagestyle{plain}

\newcommand\blfootnote[1]{%
  \begingroup
  \renewcommand\thefootnote{}\footnote{#1}%
  \addtocounter{footnote}{-1}%
  \endgroup
}
\blfootnote{$^*$ Authors ordered alphabetically.}

\begin{abstract}
A membership inference attack allows an adversary to query a trained machine learning model
to predict whether or not a particular example was contained in the model's training dataset.
These attacks are currently evaluated using average-case ``accuracy'' metrics that fail to characterize whether the attack can confidently identify any members of the training set.
We argue that attacks should instead be evaluated by computing their true-positive rate at low (e.g., $\leq 0.1\%$) false-positive rates, and find most prior attacks perform poorly when evaluated in this way.
To address this we develop a Likelihood Ratio Attack (LiRA) that carefully combines multiple ideas from the literature.
Our attack is 10$\times$ more powerful at low false-positive rates, and also
strictly dominates prior attacks on existing metrics.
\end{abstract}

\section{Introduction}

Neural networks are now trained on increasingly sensitive datasets, 
and so it is necessary to ensure that trained models are privacy-preserving.
In order to empirically verify if a model is in fact private,
membership inference attacks \cite{shokri2016membership} have become the de facto standard \cite{song2020mia,murakonda2020ml}
because of their simplicity.
A membership inference attack receives as input a trained model and
an example from the data distribution, and predicts if that
example was used to train the model.

{Unfortunately as noted by recent work \cite{murakonda2021quantifying,ye2021enhanced}, many prior membership inference attacks use an
incomplete evaluation methodology
that considers average-case success metrics (e.g., accuracy or ROC-AUC)}
that aggregate an attack's accuracy over an entire dataset and over all detection thresholds~\cite{song2020mia,sablayrolles2019white,yeom2018privacy,song2021systematic,choquette2021label,rahimian2020sampling,rahimian2020sampling,li2020membership,liu2021ml,salem2018mlleaks,nasr2018machine,pyreglis2017knock,rahman2018membership,jia2019memguard,truex2018towards, leino2019stolen, hayes2019logan}.
However, privacy is not an average case
metric, and should not be evaluated as such \cite{DPorg-average-case-dp}.
Thus, while existing membership inference attacks do appear effective when evaluated
under this average-case methodology, we make the case they do not actually
effectively measure the worst-case privacy of machine learning models.

\textbf{Contributions.}
In this paper we re-examine the problem statement of membership
inference attacks from first principles.
We first argue that membership inference attacks should be
evaluated by considering their true-positive rate (TPR)
at low false-positive rates (FPR).
This objective of designing methods around low false-positive rates
is typical in many areas of computer security \cite{lazarevic2003comparative,pantel1998spamcop,metsis2006spam,kantchelian2015better,kolter2006learning,ho2017detecting}, {and for similar reasons it is the right metric here.}
If a membership inference attack can \emph{reliably} {violate the privacy of
even just a few users in a sensitive dataset, it has succeeded.}
And conversely, an attack that only \emph{unreliably} achieves high aggregate
attack success rate should not be considered successful.

When evaluated this way, we find most prior attacks fail in the low false-positive rate regime.
{Furthermore, aggregate metrics (e.g., AUC) are often uncorrelated with low FP success rates.}
For example the attack of \citet{yeom2018privacy} has a high
accuracy ($59.5\%$) yet fails completely at low FPRs, 
and the attack of \citet{long2017towards}
has a much lower accuracy ($53.5\%$) 
but achieves {higher} success rates at low FPRs.

\begin{figure}
    \centering
    \includegraphics[scale=.75]{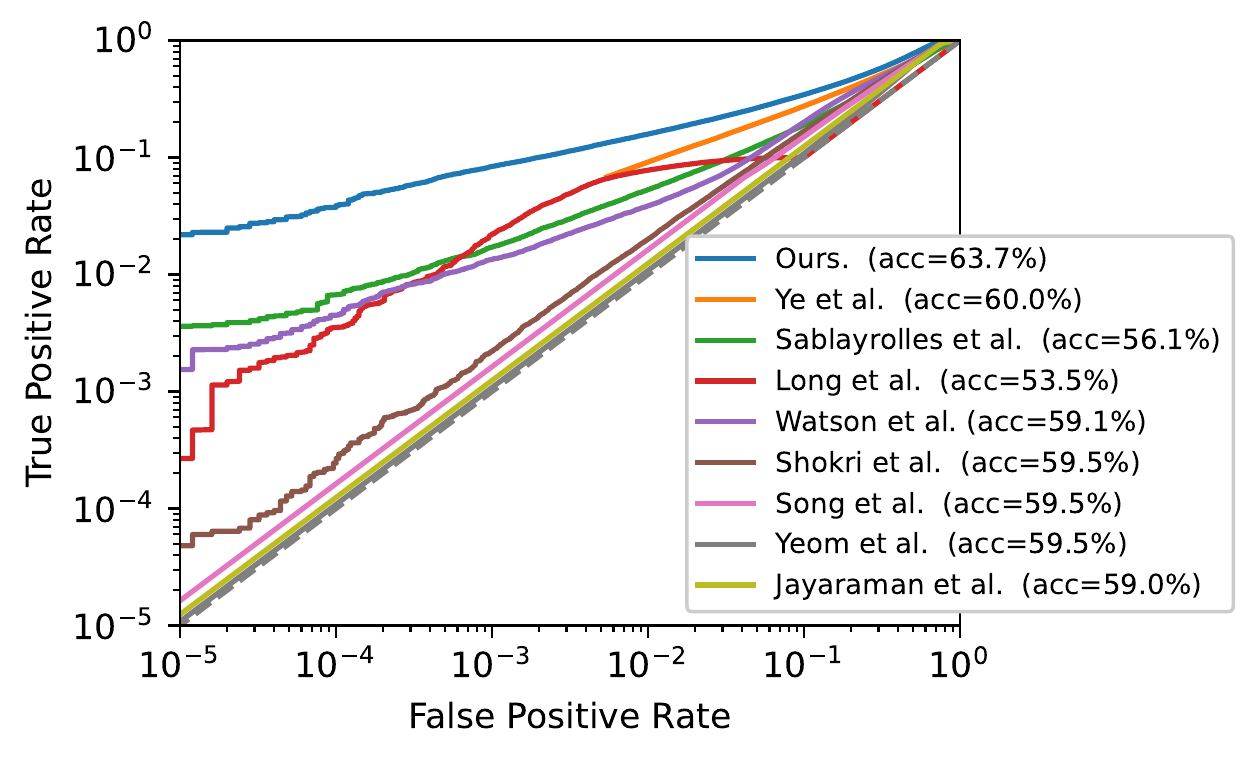}
    \vspace{-1em}
    \caption{Comparing the true-positive rate vs. false-positive rate of prior
    membership inference attacks reveals a wide gap in effectiveness. An attack's average \emph{accuracy} is not indicative of its performance at low FPRs.
    By extending on the most effective ideas, we improve
    membership inference attacks by $10\times$,
    for a non-overfit CIFAR-10 model (92\% test accuracy).
    \vspace{-1em}}
    \label{fig:full_comparison}
\end{figure}

We develop a Likelihood Ratio Attack (LiRA)
that succeeds $10\times$ more often than
prior work at low FPRs---but still strictly dominates
prior attacks on aggregate metrics introduced previously.
%
%
Our attack combines per-example difficulty scores~\cite{sablayrolles2019white, watson2021importance, long2020pragmatic} with
a principled and well-calibrated Gaussian likelihood estimate.
Figure~\ref{fig:full_comparison} shows the success rate of our attack on a log-scale
Receiver Operating Characteristic (ROC) curve \cite{sankararaman2009genomic},
comparing the ratio of true-positives to false-positives.
We perform an extensive experimental evaluation to understand 
each of the factors that contribute to our attack's success,
and release our open source code.\footnote{\url{https://github.com/tensorflow/privacy/tree/master/research/mi_lira_2021}}

Future work will need to re-examine many questions that have been studied
using prior, much less effective, membership inference attacks.
Attacks that use less information (e.g., label-only attacks~\cite{choquette2021label,rahimian2020sampling,li2020membership}) may or may
not achieve high success rate at low false-positive rates;
algorithms previously seen as ``private'' because they resist prior attacks might be vulnerable to our new attack; and
old defenses dismissed as ineffective might be able to defend against these
new stronger attacks.

\section{Background}
We begin with a background that will be familiar to readers knowledgeable of machine learning privacy.

\subsection{Machine learning notation}

A classification neural network $f_\theta : \mathcal{X} \to [0,1]^n$ is a learned function that maps some input data sample $x \in \mathcal{X}$ to an $n$-class probability
distribution; {we let $f(x)_y$ denote the probability of class $y$.}
Given a dataset $D$ sampled from some underlying distribution $\mathbb{D}$, we write
$f_\theta \gets \mathcal{T}(D)$
to denote that the neural network $f$ parameterized with weights $\theta$
is learned by running the training algorithm
$\mathcal{T}$ on the training set $D$.
Neural networks are trained via stochastic gradient descent \cite{lecun1998gradient} to minimize some loss function $\ell$:
\begin{equation}
    \theta_{i+1} \gets \theta_i - \eta \sum_{(x,y) \in B} \nabla_{\theta}\ell(f_{\theta_{i}}(x),y)
    \label{eq:grad_descent}
\end{equation}
Here, $B$ is a batch of random training examples from $D$,
and $\eta$ is the learning rate, a small constant. 
For classification tasks, the most common loss function is the cross-entropy loss:
\[
\ell(f_\theta(x),y) = - \log(f_\theta(x)_y) \;.
\]
When the weights $\theta$ are clear from context, we will simply write a trained model as $f$. 
At times it will be useful to view a model $f$ as a function $f(x) = \sigma(z(x))$, where $z: \mathcal{X} \to \mathbb{R}^n$ returns the \emph{feature outputs} of the network, followed by a \emph{softmax} normalization layer $\sigma(z) = [\frac{e^{z_1}}{\sum_{i} e^{z_i}}, \dots, \frac{e^{z_n}}{\sum_{i} e^{z_i}}]$.

Training neural networks that reach $100\%$ training accuracy is easy---running
the gradient descent from Equation~\ref{eq:grad_descent} on any sufficiently sized neural network eventually achieves this goal \cite{zhang2021understanding}.
The difficulty is in training models that generalize to an unseen
\emph{test set} $D_{\text{test}} \gets \mathbb{D}$ drawn from the same distribution.
There are a number of techniques to increase the
generalization ability of neural networks (augmentations \cite{cubuk2018autoaugment,van2001art,zhong2020random}, weight regularization \cite{krogh1992simple},
tuned learning rates \cite{jacobs1988increased,loshchilov2016sgdr}).
For the remainder of this paper, 
all models we train use state-of-the-art generalization-enhancing
techniques.
This makes our analysis much more realistic than prior work,
which often uses models with 2--5$\times$ higher error rates than our models.

\subsection{Training data privacy}

Neural networks must not leak details of their training datasets,
particularly when used in privacy-sensitive scenarios \cite{chen2019gmail,esteva2017dermatologist}.
The field of training data privacy constructs
attacks that leak data,
develops techniques to prevent memorization,
and measures the privacy of proposed defenses.

\paragraph{Privacy attacks}
There are various forms of attacks on the privacy of training data.
\emph{Training data extraction} \cite{carlini2020extracting} is an explicit attack where
an adversary recovers individual examples used to train the
model.
%
%
In contrast, \emph{model inversion} attacks recover aggregate details of particular sub-classes
instead of individual training examples 
\cite{fredrikson2015model}.
Finally, \emph{property inference} attacks aim at inferring non-trivial properties of the training dataset. For example, a classifier trained on bitcoin logs can reveal whether or not the machines that generated the logs were patched for Meltdown and Spectre~\cite{ganju2018property}.

{
We focus on a more fundamental attack that predicts if a particular example is part of a training dataset.
First explored as \emph{tracing} attacks \cite{homer2008resolving,sankararaman2009genomic,dwork2015robust,dwork2017exposed}
on medical datasets,
they were exended to machine learning models
as \emph{membership inference attacks} \cite{shokri2016membership}.
In these settings, being able to reliably (with high precision) identify a few
users as being contained in sensitive medical datasets is itself a privacy violation~\cite{homer2008resolving}---even if this is done with low recall.
Further, membership inference attacks are the foundation of stronger extraction attacks \cite{carlini2019secret, carlini2020extracting}, and in order to be used in this way must again have exceptionally high precision.
}

\paragraph{Theory of memorization}
%
The ability to perform membership inference is directly tied to a model's ability to \emph{memorize} individual data points or labels. \citet{zhang2021understanding} demonstrated that standard neural networks can memorize entirely randomly labeled datasets. A recent line of work initiated by 
\citet{feldman2020does} shows both theoretically and empirically that some amount of memorization may be \emph{necessary} to achieve optimal generalization \cite{feldman2020neural, brown2021memorization}. 

\paragraph{Privacy-preserving training}
%
The most widely deployed technique to make neural networks private is to
make the learning process \emph{differentially private} \cite{dwork2014algorithmic}.
This can be done in various ways---for example by modifying the SGD algorithm \cite{song2013stochastic,abadi2016deep}, or by aggregating results from a model ensemble \cite{papernot2018scalable}.
Independent from differential privacy based defenses, there are other
heuristic techniques (that is, without a formal proof of privacy) 
that have been developed to improve the privacy of
machine learning models \cite{nasr2018machine,jia2019memguard}.
Unfortunately, many of these have been shown to be vulnerable to more advanced
forms of attack \cite{choquette2021label,song2021systematic}.

\paragraph{Measuring training data privacy}
Given a particular training scheme, a final direction of work aims to
answer the question ``how much privacy does this scheme offer?''
Existing techniques often work by altering the training pipeline, 
either by injecting outlier canaries \cite{carlini2019secret},
or using poisoning to search for worst-case memorization \cite{jagielski2020auditing,nasr2021adversary}.
While these techniques give increasingly strong measurements of a trained
model's privacy, the fact that they require modifying the training pipeline
creates an up-front cost to deployment.
As a result,
by far the most common technique used to audit machine learning
models is to just use a membership inference attack.
Existing 
membership inference attack libraries (see, e.g., \citet{song2020mia,murakonda2020ml})
form the basis for most production privacy analysis \cite{song2020mia},
and it is therefore critical that they accurately assess the privacy of
machine learning models.

%

\section{Membership Inference Attacks}
The objective of a membership inference attack (MIA) \cite{shokri2016membership} 
is to predict if a specific
training example
was, or was not, used as training data in a particular model.
This makes MIAs the simplest and most widely deployed attack for auditing training data privacy. It is thus important that they can reliably succeed at this task.
%
%
This section formalizes the membership inference attack security game (\S\ref{ssec:def}),
and introduces our membership inference evaluation methodology (\S\ref{ssec:eval}).

\subsection{Definitions}
\label{ssec:def}
\newtheorem{definition}{Definition}

We define membership inference via a standard security game inspired by~\citet{yeom2018privacy} and \citet{jayaraman2020revisiting}.

\begin{definition}[Membership inference security game]
\label{def:game}
The game proceeds between a challenger $\mathcal{C}$ and an adversary $\mathcal{A}$:
\begin{enumerate}
    \item The challenger samples a training dataset $D \gets \mathbb{D}$ and trains a model $f_\theta \gets \mathcal{T}(D)$ on the dataset $D$.
    \item The challenger flips a bit $b$, and if $b=0$, samples a fresh challenge point from the distribution $(x, y) \gets \mathbb{D}$ {(such that $(x, y) \notin D$)}. Otherwise, the challenger selects a point from the training set $(x, y) \gets^{\$} D$.
    \item The challenger sends $(x, y)$ to the adversary.
    \item The adversary gets query access to the distribution $\mathbb{D}$, and to the model $f_\theta$, and outputs a bit $\hat{b} \gets \mathcal{A}^{\mathbb{D}, f}(x, y)$.
    \item Output 1 if $\hat{b}=b$, and 0 otherwise.
\end{enumerate}
\end{definition}
For simplicity, we will write $\mathcal{A}(x, y)$ to denote the adversary's prediction on the sample $(x,y)$ when the distribution $\mathbb{D}$ and model $f$ are clear from context.

Note that this game assumes that the adversary is given access to the underlying training
data distribution  $\mathbb{D}$;
while some attacks do not make use of this assumption~\cite{yeom2018privacy},
many attacks require query-access to the distribution in order to train ``shadow models''~\cite{shokri2016membership}
 (as we will describe).
The above game also assumes that the adversary is given access to both a training
example \emph{and} its ground-truth label.

Instead of outputting a ``hard prediction'', all the attacks we consider output 
a continuous \emph{confidence score}, which is then thresholded to yield a membership prediction.
That is, 
\[\mathcal{A}(x, y) = \mathbbm{1}[\mathcal{A}'(x, y) > \tau]
\]
where $\mathbbm{1}$ is the indicator function, $\tau$ is some tunable decision threshold, and $\mathcal{A}'$ outputs a real-valued confidence score.

%

\bigskip \noindent \textbf{A first membership inference attack.}
For illustrative purposes, we begin by considering a very simple membership inference attack (due to~\citet{yeom2018privacy}).
This attack relies on the observation that, because machine learning models are
trained to minimize the loss of their training examples (see Equation~\ref{eq:grad_descent}),
 examples with lower loss are on average more likely to be members of the training data.
Formally, the LOSS membership inference attack defines
\[\mathcal{A}_{\text{loss}}(x,y) = \mathbbm{1}[-\ell(f(x), y) > \tau] \;. \]

\subsection{Evaluating membership inference attacks}
\label{ssec:eval}

Prior work lays out several strategies to 
determine the effectiveness of a membership inference attack, i.e., how to measure the adversary's success in \Cref{def:game}.
We now show that existing evaluation methodologies fail to characterize 
whether an attack succeeds at confidently predicting membership. 
We thus propose a more suitable evaluation procedure.

As a running example for the remainder of this section, we train a standard CIFAR-10~\cite{cifar}
ResNet~\cite{he2015deep} to 92\% test accuracy by training it on half of the dataset
(i.e., $25{,}000$ examples)---%
leaving another $25{,}000$ examples for evaluation as non-members.
While this dataset is not \emph{sensitive}, it serves as a strong baseline for understanding
properties of machine learning models in general.
We train this model using standard techniques to reduce overfitting, including 
weight decay \cite{krogh1992simple}, train-time augmentations \cite{cubuk2018autoaugment}, and early stopping.
As a result, this model has only a $8\%$ train-test accuracy gap.

\bigskip \noindent \textbf{Balanced Attack Accuracy.}
The simplest method to evaluate attack efficacy is through 
a standard ``accuracy'' metric that measures how often an attack
correctly predicts membership on a balanced dataset of members and non-members~\cite{shokri2016membership, yeom2018privacy, sablayrolles2019white,nasr2019comprehensive, song2021systematic, choquette2021label, truex2018towards, leino2019stolen, hayes2019logan, watson2021importance}.

\begin{definition}
The \emph{balanced attack accuracy} of a membership inference attack $\mathcal{A}$ in Definition~\ref{def:game}
is defined as
\[\Pr_{x,y,f,b}[\mathcal{A}^{\mathbb{D},f}(x, y) = b]. \]
\end{definition}


Even though balanced accuracy is used in many papers to evaluate membership inference attacks, we argue that this metric is inherently inadequate for multiple reasons:

\begin{itemize}
    \item Balanced accuracy is \emph{symmetric}. That is, the metric assigns equal cost to false-positives and to false-negatives.
    However, in practice, adversaries often only care about one of these two sources of errors.
    For example, when a membership inference attack is used in a training data extraction attack~\cite{carlini2020extracting}, false negatives are benign (some data will not be successfully extracted) whereas false-positives directly reduce the utility of the attack.
    \item Balanced accuracy is an \emph{average-case} metric, but this is not what matters in security. Consider comparing two attacks. Attack A perfectly targets a known subset of $0.1\%$ of users, but succeeds with a random $50\%$ chance on the rest. Attack B succeeds with $50.05\%$ probability on any given user. On average, these two attacks have the same attack success rate (and thus the same balanced accuracy). However, the second attack is practically useless, while the first attack is exceptionally potent. 
    
\end{itemize}

We now illustrate how exactly these issues arise for the simple LOSS attack described above. For our CIFAR-10 model, this attack's balanced accuracy is $60\%$. This is (much) better than random guessing, and so one might reasonably conclude that the attack is useful and practically worrying.

However, this attack completely fails at \emph{confidently} identifying \emph{any} members! Let's examine for the moment the $1\%$ of samples from the CIFAR-10 dataset with lowest losses $\ell(f(x), y)$. These are the samples where the attack is most confident that they are members. Yet, on this subset, the attack is only correct $48\%$ of the time (\emph{worse} than random guessing). In contrast, for the 1\% samples with highest loss (confident non-members), the attack is correct $100\%$ of the time.
\textbf{Thus, the LOSS attack is actually a strong \emph{non-membership} inference attack}, and is practically useless at inferring membership.
An attack with the symmetrical property (i.e., the attack confidently identifies members, but not non-members) is a much stronger attack on privacy, yet it achieves the same balanced accuracy.

\bigskip \noindent \textbf{ROC Analysis.}
Instead of the balanced accuracy, we should thus consider metrics that emphasize positive predictions (i.e., membership guesses) over negative (non-membership) predictions.
A natural choice is to consider the tradeoff between the true-positive rate (TPR) and false-positive rate (FPR). Intuitively, an attack should maximize the true-positive rate (many members are identified), while incurring few false-positives (incorrect membership guesses).
We prefer this to a precision/recall analysis because TPR/FPR is independent of the (often unknown) prevalence of members in the population.

\begin{figure}[h]
    \centering
    
    \begin{subfigure}[b]{0.48\columnwidth}
    \centering
    \includegraphics[width=\textwidth]{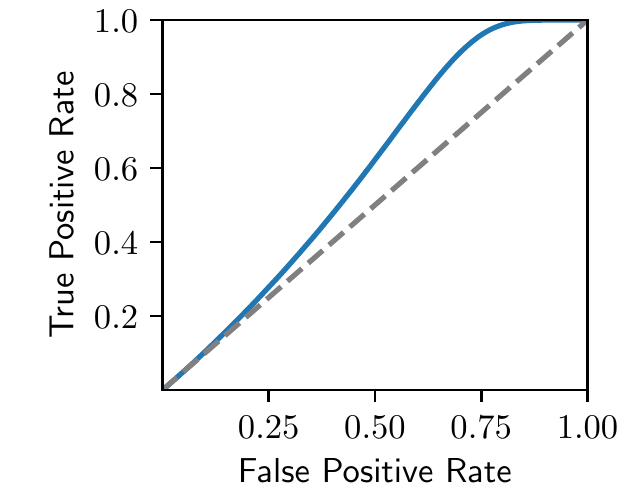}
    \caption{linear scale}
    \label{fig:roc_curve_linear}
    \end{subfigure}
    \begin{subfigure}[b]{0.48\columnwidth}
    \centering
    \includegraphics[width=\textwidth]{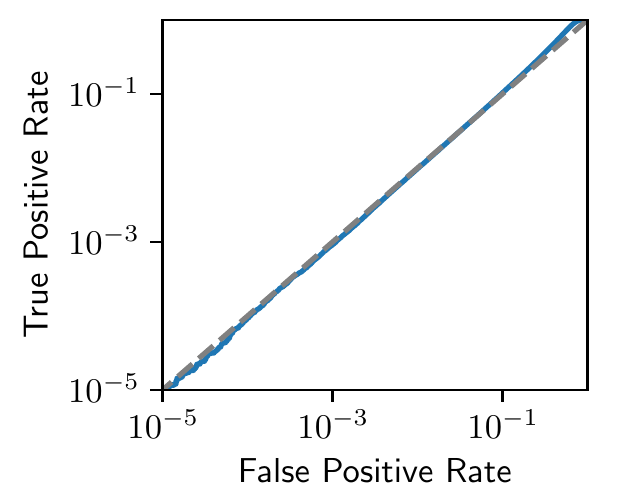}
    \caption{log scale}
    \label{fig:roc_curve_log}
    \end{subfigure}
    \caption{ROC curve for the LOSS baseline membership inference attack, shown with both linear scaling (left), also and log-log scaling (right) to emphasize the low-FPR regime.}
\end{figure}

The TPR/FPR tradeoff is fully characterized by the Receiver Operating Characteristic (ROC) curve, which compares the attack's TPR and FPR for all possible choices of the decision threshold $\tau$. In \Cref{fig:roc_curve_linear}, we show the ROC curve for the LOSS attack. The attack fails to achieve a TPR better than random chance at any FPR below $20\%$---it is therefore ineffective at confidently breaching the privacy of its members.

Prior papers that do report ROC curves summarize them by the AUC (Area Under the Curve)~\cite{salem2018mlleaks, melis2019exploiting, he2021stealing, watson2021importance, ye2021enhanced, murakonda2019ultimate}.
However, as we can see from the curves above, the AUC is not an appropriate measure of an attack's efficacy, since the AUC averages over all false-positive rates, including high error rates that are irrelevant for a practical attack.
The TPR of an attack when the FPR is above 50\% is not meaningfully useful, yet this regime accounts for more than half of its AUC score.

To illustrate, consider our hypothetical Attack A from earlier that confidently identifies 0.1\% of members, but makes no confident predictions for any other samples. This attack perfectly breaches the privacy of some members, but has an AUC $\approx 51\%$---lower than the AUC of the weak LOSS attack.

\smallskip \noindent \textbf{True-Positive Rate at Low False-Positive Rates.}
Our recommended evaluation of membership inference attacks is thus to report an attack's true-positive rate at \emph{low} false-positive rates.

Prior work occasionally reports true-positive rates at moderate false-positive rates (or reports precision/recall values that can be converted into TPR/FPR rates if the prevalence is known).
For example, \citet{shokri2016membership} frequently reports that the 
``recall is almost 1''
however there is a meaningful FPR difference between a recall of 1.0 and 0.999.
Other works consistently report precision/recall values, but for equivalent false-positive rates between $3\%$ and $40\%$, which we argue is too high
to be practically meaningful.

In this paper, we argue for studying the extremely low false-positive regime. We do this by (1) reporting full ROC curves in logarithmic scale (see \Cref{fig:roc_curve_log}); and (2) optionally summarizing an attack's success rate by reporting its TPR at a fixed low FPR (e.g., $0.001\%$ or $0.1\%$).
For example, the LOSS attack achieves a TPR of $0\%$ at an FPR of $0.1\%$ (worse than chance).
While summarizing an attack's performance at a single choice of (low) FPR can be useful for quickly comparing attack configurations, we encourage future work to always also report full (log-scale) ROC curves as we do.

\section{The Likelihood Ratio Attack (LiRA)}
\label{sec:bmia}


\subsection{Membership inference as hypothesis testing}

The game in \Cref{def:game} requires the adversary to distinguish between two ``worlds'': one where $f$ is trained on a randomly sampled dataset that contains a target point $(x,y)$, and one where $f$ is not trained on $(x, y)$. 
{
It is thus natural to see a membership inference attack as performing a \emph{hypothesis test} to guess whether or not $f$ was trained on $(x, y)$.
}

We formalize this by considering two distributions over models: $\mathbb{Q}_{\text{in}}(x, y) = \{f \gets \mathcal{T}(D \cup \{(x,y)\}) \mid D \gets \mathbb{D}\}$ is the distribution of models trained on datasets containing $(x, y)$, and then $\mathbb{Q}_{\text{out}}(x, y) = \{f \gets \mathcal{T}(D {\setminus \{(x,y)\}}) \mid D \gets \mathbb{D}\}$.
{
Given a model $f$ and a target example $(x, y)$, 
the adversary's task is to perform a hypothesis test that
predicts if $f$ was sampled either from $\mathbb{Q}_{\text{in}}$ or if it was sampled from $\mathbb{Q}_{\text{out}}$ \cite{sankararaman2009genomic}.}

{We perform this hypothesis test according to the}
Neyman-Pearson lemma \cite{neyman1933ix}, which states that the best hypothesis test at a fixed false positive rate is obtained by thresholding the \emph{Likelihood-ratio Test} between the two hypotheses:
\begin{equation}
    \label{eq:likelihood_test}
    \Lambda(f; x,y) = \frac{p(f \mid \mathbb{Q}_{\text{in}}(x, y))}{p(f \mid \mathbb{Q}_{\text{out}}(x, y))} \;,
\end{equation}
where $p(f \mid \mathbb{Q}_b(x, y))$ is the probability density function over $f$ under the (fixed) distribution of model parameters $\mathbb{Q}_b(x, y)$.

Unfortunately the above test is intractable: even the distributions $\mathbb{Q}_{\text{in}}$ and $\mathbb{Q}_{\text{out}}$ are not analytically known. 
{
To simplify the situation, we instead define 
$\tilde{\mathbb{Q}}_{\text{in}}$ and $\tilde{\mathbb{Q}}_{\text{out}}$ as the distributions of \emph{losses} on $(x,y)$ for models either trained, or not trained, on this example.}
Then, we can replace both probabilities in Equation \ref{eq:likelihood_test} with the easy-to-calculate quantity
\begin{equation}
    p(\ell(f(x), y) \mid \tilde{\mathbb{Q}}_{\text{in/out}}(x, y)).
    \label{eqn:simpler}
\end{equation}
This is now a likelihood test for a one-dimensional statistic, which can be efficiently computed with query access to $f$.

\smallskip \noindent
{\textbf{Our attack} follows the above intuition.
We train several ``shadow models'' in order to directly estimate the distribution
$\tilde{\mathbb{Q}}_{\text{in/out}}$.
To minimize the number of shadow models necessary, we 
assume $\tilde{\mathbb{Q}}_{\text{in/out}}$ is a Gaussian distribution,
reducing our attack to estimating just four parameters: the mean and variance of each distribution.
To run our inference attack on any model $f$, we can compute its loss on $\ell(f(x), y)$, measure the likelihood of this loss under each of the distributions $\tilde{\mathbb{Q}}_{\text{in}}$ and $\tilde{\mathbb{Q}}_{\text{out}}$, and return whichever is is more likely.
}

\subsection{Memorization and per-example hardness}

By casting membership inference as a Likelihood-ratio test, it becomes clear why the LOSS attack (and those that build on it) are ineffective:
by directly thresholding the quantity $\ell(f(x), y)$, this attack implicitly assumes that the losses of all examples are a priori on an equal scale, and that the inclusion or exclusion of one example will have a similar effect on the model as any other example.
That is, if we measure $\ell(f(x), y) < \ell(f(x'), y')$ 
then the LOSS attack predicts that $(x, y)$ is more likely to be a member than $(x', y')$---regardless 
of any other properties of these examples.

\citet{feldman2020neural} show that not all examples are equal: some examples (``outliers'') have an outsized effect on
the learned model when inserted into a training dataset, compared to other (``inlier'') examples.
To replicate their experiment, we choose a training dataset $D$
and sample a random subset $D_{\text{in}} \subset D$ containing half of the dataset.
We train a model on this dataset $f \gets \mathcal{T}(D_{\text{in}})$, and evaluate the loss on every example $(x, y) \in D$, annotated by whether or not $(x, y)$ was in the training set $D_{\text{in}}$.
We repeat the above experiment hundreds of times, 
thereby empirically estimating the distributions 
$p(\ell(f(x), y) \mid \tilde{\mathbb{Q}}_{\text{in/out}}(x, y))$ by sampling.

\begin{figure}[t]
    \centering
    \includegraphics[scale=0.75]{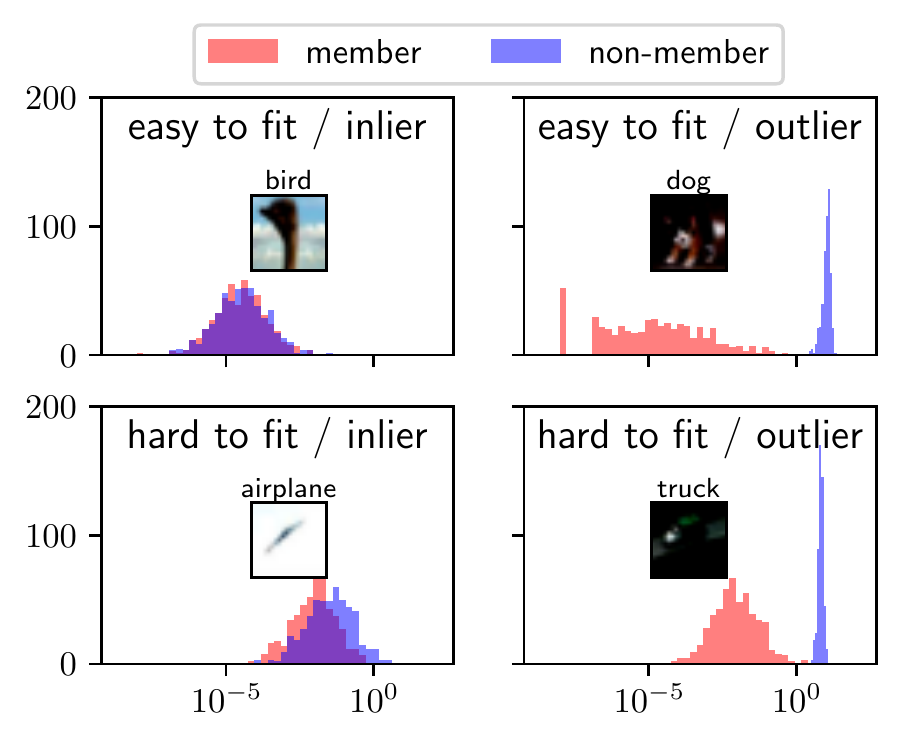}
    \caption{Some examples are easier to fit than others, and some have a larger separability between their losses when being a member of the training set or not. We train 1024 models on random subsets of CIFAR-10 and plot the losses for four examples when the example is a member of the training set ($\tilde{\mathbb{Q}}_{\text{in}}(x,y)$, in red) or not ($\tilde{\mathbb{Q}}_{\text{out}}(x,y)$, in blue).}
    \label{fig:quadrant}
\end{figure}

Figure~\ref{fig:quadrant} plots histograms of model losses on
four CIFAR-10 images when the image is contained in the
model's training dataset (red) and when it is absent (blue).
We chose these images to illustrate two different axes of variation. On the columns we compare ``inliers'' to ``outliers'', as determined by the model's loss when not trained on the example. The left column shows examples with low loss when omitted from the training set, those in the right column have high loss.
On rows we compare how easy the examples are to fit. The examples in the top row have very low loss when trained on, while the examples in the bottom row have higher loss.
Importantly, observe that these two dimensions do measure different quantities.
An example can be an outlier but easy to fit (upper right), or an inlier but hard to fit (lower left).

The goal of a membership inference adversary is to distinguish the two distributions in Figure~\ref{fig:quadrant} for a given example. 
This view illustrates the shortcomings of prior attacks (e.g., the LOSS attack): a global threshold on the observed loss $\ell(f(x), y)$ cannot distinguish between the different scenarios in Figure~\ref{fig:quadrant}. The only confident assessment that such an attack can make is that examples with high loss are \emph{non}-members.
In contrast, the Likelihood-ratio test in \Cref{eq:likelihood_test} considers the hardness of each example individually by modeling separate pairs of distributions $\tilde{\mathbb{Q}}_{\text{in}}, \tilde{\mathbb{Q}}_{\text{out}}$ for each example $(x,y)$.

\subsection{Estimating the likelihood-ratio {with parametric modeling}}
\label{ssec:estimating_likelihood}

We directly turn this observation into a membership inference attack
by computing \emph{per-example hardness scores}~\cite{sablayrolles2019white, long2020pragmatic, watson2021importance, ye2021enhanced}.
By training models on random samples of data from the distribution $\mathbb{D}$, we obtain empirical estimates of the distributions $\tilde{\mathbb{Q}}_{\text{in}}$ and $\tilde{\mathbb{Q}}_{\text{out}}$ for any example $(x, y)$. 
And from here, we can estimate the likelihood from Equation~\ref{eqn:simpler} to predict if an example is a member of the training dataset or not.

\begin{figure}[t]
    \centering
    \vspace{-1em}
    \includegraphics[width=\columnwidth]{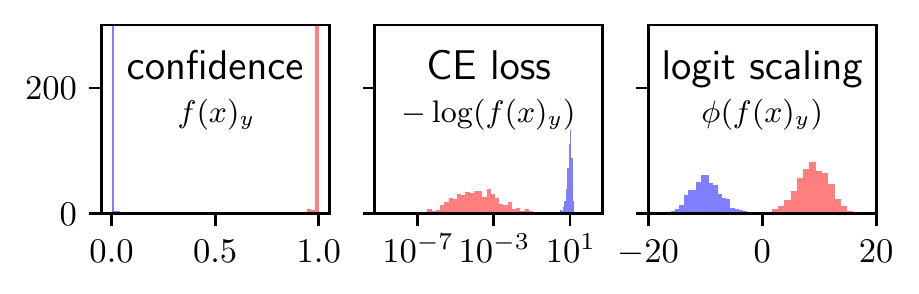}
    \caption{The model's confidence, or its logarithm (the cross-entropy loss) are not normally distributed. Applying the logit function yields values that are approximately normal.}
    \label{fig:logits}
\end{figure}




To improve performance at very low false-positive rates, instead of empirically modeling the distributions $\tilde{\mathbb{Q}}_{\text{in/out}}$ 
directly from the data, we opt for a \emph{parametric} 
and model $\tilde{\mathbb{Q}}_{\text{in/out}}$ by Gaussian distributions.
{Parametric modeling has several significant benefits over nonparametric modeling.}
\begin{itemize}
    \item {Parametric modeling requires training fewer shadow models to achieve the same generalization
    of nonparametric approaches. For example, we can match the recent (nonparametric) work of \cite{ye2021enhanced} with 400$\times$ fewer models.}
    \item {We can extend our attack to multivariate parametric models, allowing us to further improve attack success rate by querying the model multiple times (\S\ref{ssec:num_queries}).}
\end{itemize}

Doing this requires some care.
Indeed, as can be seen in \Cref{fig:quadrant}, the model's cross-entropy loss is \emph{not} well approximated by a normal distribution. First, the cross-entropy loss is on a logarithmic scale. If we take the negative exponent, $\exp(-\ell(f(x), y))$, we instead obtain the model ``confidence'' $f(x)_y$, which is bounded in the interval $[0, 1]$ and thus not normally distributed either (i.e., the confidences for outliers and inliers concentrate, respectively, around $0$ and $1$).
We thus apply a \emph{logit} scaling to the model's confidence,
\[
\phi(p) = \log\left(\frac{p}{1-p}\right), \quad \text{ for } p=f(x)_y
\]
to obtain a statistic in the range $(-\infty, \infty)$ that is (empirically) approximately normal.
\Cref{fig:logits} displays the distributions of model confidences, the negative log of the confidences (the cross-entropy loss), and the logit of the confidences. Only the logit appraoch is well approximated by a pair of Gaussians.

\bigskip \noindent
\textbf{Our complete online attack (\Cref{alg:main}).}
We first train $N$ shadow models~\cite{shokri2016membership} on random samples from the data distribution $\mathbb{D}$, so that half of these models are trained on the target point $(x,y)$, and half are not (we call these respectively IN and OUT models for $(x,y)$).
We then fit two Gaussians to the confidences of the IN and OUT models on $(x,y)$ (in logit scale). Finally, we query the confidence of the target model $f$ on $(x, y)$ and output a parametric Likelihood-ratio test.
%

This attack is easily parallelized across multiple target points. Given a dataset $D \gets \mathbb{D}$, we train shadow models on $N$ subsets of $D$, chosen so that each target $(x,y) \in D$ appears in $N/2$ subsets. The same $N$ shadow models can then be used to estimate the Likelihood-ratio test for all examples in $D$.

As an optimization, we can improve the attack by querying the target model on multiple points $x_1, x_2, \dots, x_m$ obtained by applying standard data augmentations to the target point $x$ (as previously observed in~\cite{choquette2021label}). In this case, we fit $m$-dimensional spherical Gaussians $\mathcal{N}(\boldsymbol{\mu}_{\text{in}}, \boldsymbol{\sigma}^2_{\text{in}}I), \mathcal{N}(\boldsymbol{\mu}_{\text{out}}, \boldsymbol{\sigma}^2_{\text{out}}I)$ to the losses collected from querying the shadow models $m$ times per example, and compute a standard likelihood-ratio test between two multivariate normal distributions.

\begin{algorithm}[t]
 \begin{algorithmic}[1]
  \REQUIRE \text{model} $f$, \text{example} $(x, y)$, \text{data distribution} $\mathbb{D}$
  \STATE $\text{confs}_{\text{in}} = \{\}$
  \STATE $\text{confs}_{\text{out}} = \{\}$
  \FOR{$N$ times}
    \STATE $D_{\text{attack}} \gets^\$ \mathbb{D}$ \algcomment{Sample a shadow dataset}
    \STATE $f_{\text{in}} \gets \mathcal{T}(D_{\text{attack}} \cup \{(x,y)\})$ \algcomment{train IN model}
    \STATE $\text{confs}_{\text{in}} \gets \text{confs}_{\text{in}} \cup \{\phi(f_{\text{in}}(x)_y)\}$
    \STATE $f_{\text{out}} \gets \mathcal{T}(D_{\text{attack}} {\setminus \{(x,y)\}})$ \algcomment{train OUT model}
    \STATE $\text{confs}_{\text{out}} \gets \text{confs}_{\text{out}} \cup \{\phi(f_{\text{out}}(x)_y)\}$
  \ENDFOR
  \STATE $\mu_{\text{in}} \gets \texttt{mean}(\text{confs}_{\text{in}})$
  \STATE $\mu_{\text{out}} \gets \texttt{mean}(\text{confs}_{\text{out}})$
  \STATE $\sigma_{\text{in}}^2 \gets \texttt{var}(\text{confs}_{\text{in}})$
  \STATE $\sigma_{\text{out}}^2 \gets \texttt{var}(\text{confs}_{\text{out}})$
  \STATE $\text{conf}_{\text{obs}} = \phi(f(x)_y)$ \algcomment{query target model}
  \vspace{0.5em}
  \RETURN $\displaystyle \Lambda = \frac{p(\text{conf}_{\text{obs}}\ \mid\ \mathcal{N}(\mu_{\text{in}}, \sigma^2_{\text{in}}))}
    { p(\text{conf}_{\text{obs}}\ \mid\ \mathcal{N}(\mu_{\text{out}}, \sigma^2_{\text{out}}))}$
 \end{algorithmic}
 \caption{\textbf{Our online Likelihood Ratio Attack (LiRA).}
 We train shadow models on datasets with and without the target example,
 estimate mean and variance of the loss distributions, and compute
 a likelihood ratio test.
 (In our \textbf{offline} variant, we omit lines 5, 6, 10, and 12,
 and instead return the prediction by estimating a single-tailed distribution, as is shown in \Cref{eq:offline_test}.)
 }
 \label{alg:main}
\end{algorithm}

\bigskip \noindent
\textbf{Our offline attack.}
While our online attack is effective, it
has a significant usability limitation: it requires the adversary
train new models \emph{after} they are told to infer the membership of the example $(x,y)$.
This requires training new machine learning models for every (batch of) membership inference queries, and is computationally expensive.

To improve the efficiency of our attack, we propose an \emph{offline} attack
algorithm that trains shadow models on randomly sampled datasets ahead of time, and never trains shadow models on the target points.
For this attack, we remove lines 5,6,10 and 12 from \Cref{alg:main}, and only estimate the mean $\mu_{\text{out}}$ and variance $\sigma_{\text{out}}^2$ of model confidences when the target example is \emph{not} in the shadow models' training data.
We then change the likelihood-ratio test in line 15 to a one-sided hypothesis test. 
That is, we measure the probability of observing a confidence as high as the target model's under the null-hypothesis that the target point $(x,y)$ is a non-member:
\begin{equation}
\label{eq:offline_test}
\Lambda = 1-\Pr[Z > \phi(f(x)_y)], \text{where } Z \sim \mathcal{N}(\mu_{\text{out}}, \sigma^2_{\text{out}}) \;.
\end{equation}
The larger the target model's confidence is compared to $\mu_{\text{out}}$, the higher the likelihood that the query sample is a member.
Similar to our online attack, we improve the attack by querying on multiple augmentations and fitting a multivariate normal.

\section{Attack Evaluation}
\label{sec:eval}

We now investigate our offline and online attack variants
in a thorough evaluation across datasets and ML techniques.

{Again, we focus extensively on the low-false positive rate
regime.
This is the setting with the most practical consequences:
for example, to extract training data \cite{carlini2020extracting} it is far more important for attacks to have a low false positive rate than high average success,
as false positives are fare more costly than false negatives.
Similarly, de-identifying even a few users contained in a sensitive dataset
is far more important than saying an average-case statement 
``most people are probably not contained in the sensitive dataset''.
}

We use both datasets 
traditionally used for membership inference attack evaluations,
but also new datasets that are less typically used.
In addition to the CIFAR-10 dataset introduced previously, we also consider
three other datasets: CIFAR-100~\cite{cifar} (another standard image
classification task), ImageNet~\cite{deng2009imagenet} (a standard challenging image classification task) and WikiText-103~\cite{merity2016pointer} (a natural language processing text dataset).
For CIFAR-100, we follow the same process as for CIFAR-10 and train a wide ResNet~\cite{zagoruyko2016wide} to 60\% accuracy on half of the dataset (25{,}000 examples).
For ImageNet, we train a ResNet-50 on 50\% of the dataset (roughly half a million examples).
For WikiText-103, we use the GPT-2 tokenizer~\cite{radford2019language} to split the dataset into a million sentences 
and train a small GPT-2 \cite{radford2019language} model on 50\% of the dataset for 20 epochs to minimize the cross-entropy loss.
Prior work has additionally performed experiments on two toy datasets 
that we do not believe are meaningful benchmarks for privacy because of
their simplicity: Purchase and Texas (see~\cite{shokri2016membership} for details).%
\footnote{While these datasets ostensibly have privacy-relevance, we
believe it is more important to study datasets that reveal interesting
properties of machine learning than datasets that discuss privacy.
We nevertheless present these results in the Appendix, but encourage
future work to omit these results and focus on the more informative
tasks we consider.}

For each dataset, the adversary trains $N$ shadow models ($N=64$ for ImageNet, and $N=256$ otherwise) on training sets chosen so that each example $(x, y)$ is contained in exactly half of the shadow models' training sets (thus, for each example we have $N/2$ IN models, and $N/2$ OUT models).
We use the entire dataset for this purpose, and thus the training sets of individual shadow models and the target model may partially overlap. This is a strong assumption, which we make here mainly due to the small size of some of the datasets we consider. In \Cref{ssec:mismatched_data}, we show that our attack works just as well when the adversary trains shadow models on datasets that are fully disjoint from the target model's training set.

For all datasets except ImageNet, we repeat each attack 10 times and report the attack success rates across all 10 attacks.

\begin{figure}[t]
    \centering
    \includegraphics[scale=.75]{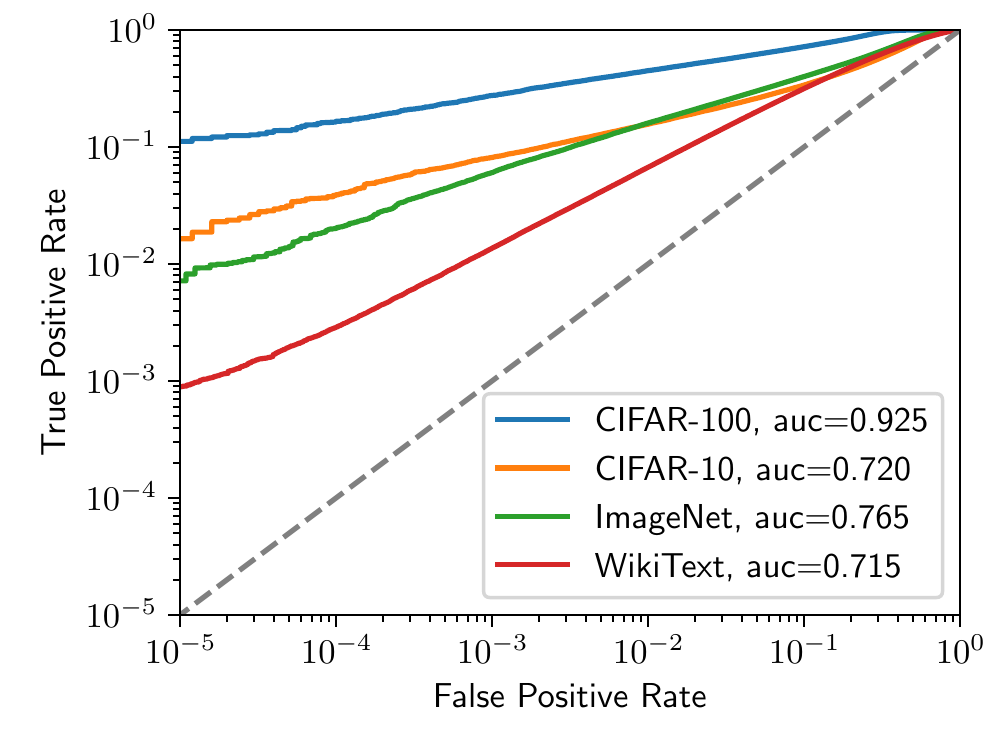}
    \caption{\textbf{Success rate of our attack on CIFAR-10, CIFAR-100, ImageNet, and WikiText.} 
    All plots are generated with 256 shadow models, except ImageNet which uses 64.
    }
    \label{fig:main}
\end{figure}

\subsection{Online attack evaluation}

\Cref{fig:main} presents the main results of our online attack when
evaluated on the four more complex of the datasets mentioned above
(CIFAR-10, CIFAR-100, ImageNet, and WikiText-103).
%
%
Even though these datasets are complex, it is relatively efficient to
train most of these models---for example a CIFAR-10 or CIFAR-100 model takes just six minutes to train.
Additional results for the Purchase and Texas dataset are given
in the Appendix---these datasets are much simpler and while
they are typically used for membership inference, we argue they are
too simple to have generalizable lessons.

Our attack has true-positive rates ranging from $0.1\%$ to $10\%$
at a false-positive rate of $0.001\%$.
If we compare the three image datasets, consistent with prior works,
we find that the
attack's \emph{average} success rate (i.e., the AUC) is correlated directly with the generalization
gap of the trained model.
All three models have perfect $100\%$ training accuracy, but the
test accuracy of the CIFAR-10 model is 90\%,
the ImageNet model is 65\%, and the CIFAR-100 model is 60\%. Yet, at low false-positives, the CIFAR-10 models are easier to attack than the ImageNet models, despite their better generalization.

\begin{figure}
    \centering
    \includegraphics[scale=.75]{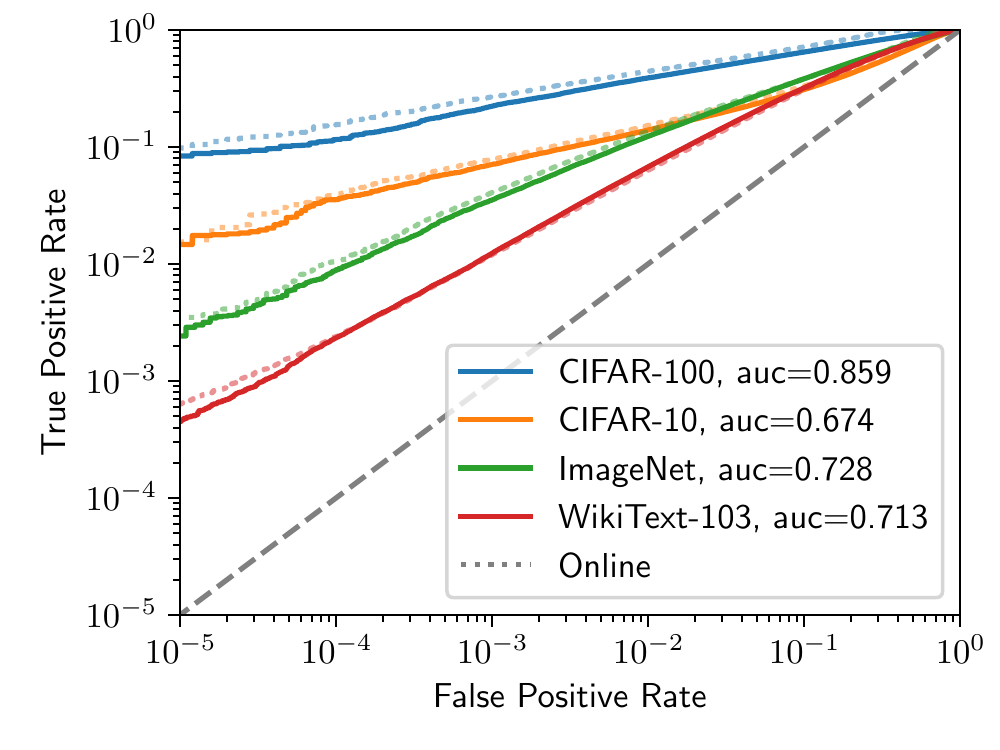}
    \caption{\textbf{Success rate of our offline attack on CIFAR-10, CIFAR-100, ImageNet, and WikiText.} 
    All plots are generated with 128 OUT shadow models, except ImageNet which uses 32. For each dataset, we also plot our online attack with the same number of shadow models (half IN, half OUT).}
    \label{fig:noin}
\end{figure}

\begin{table*}[]
    \small
    \centering
    \begin{tabular}{@{} l@{\hskip 5pt} cccc rrr rrr rrr @{}}
         & \multirow{3}{*}{\rotatebox{90}{\parbox[c]{1.2cm}{\footnotesize{shadow models}}}}
         & \multirow{3}{*}{\rotatebox{90}{\parbox[c]{1.2cm}{\footnotesize{multiple queries}}}}
         & \multirow{3}{*}{\rotatebox{90}{\parbox[c]{1.2cm}{\footnotesize{class \\hardness}}}}
         & \multirow{3}{*}{\rotatebox{90}{\parbox[c]{1.2cm}{\footnotesize{example hardness}}}}
         & \\[0.3em]
         &&&&&\multicolumn{3}{c}{TPR @ 0.001\% FPR}  & \multicolumn{3}{c}{TPR @ 0.1\% FPR} &  \multicolumn{3}{c}{Balanced Accuracy}  \\
        \cmidrule(l{5pt}r{5pt}){6-8}\cmidrule(l{5pt}r{5pt}){9-11}\cmidrule(l{5pt}r{0pt}){12-14}
        Method & 
        &&&&
        C-10 & C-100 & WT103 & C-10 & C-100 & WT103 &  C-10 & C-100 & WT103 \\
        \midrule
        \citet{yeom2018privacy} &\Circle&\Circle&\Circle&\Circle &0.0\%&0.0\%& 0.00\% & 0.0\% & 0.0\% & 0.1\%   & 59.4\% & 78.0\% & 50.0\% \\
        \citet{shokri2016membership} &\CIRCLE&\Circle&\CIRCLE&\Circle &0.0\%&0.0\%& -- & 0.3\% & 1.6\%  & --    & 59.6\% & 74.5\% & --\\
        \citet{jayaraman2020revisiting} &\Circle&\CIRCLE&\Circle&\Circle &0.0\%&0.0\%& -- & 0.0\% & 0.0\% & -- &  59.4\% & 76.9\% & --\\
        \citet{song2021systematic} &\CIRCLE&\Circle&\CIRCLE&\Circle &0.0\%&0.0\%& -- & 0.1\% & 1.4\% & --    & 59.5\% & 77.3\% & --\\
        \citet{sablayrolles2019white} &\CIRCLE&\Circle&\CIRCLE&\CIRCLE &0.1\% &0.8\%& 0.01\% & 1.7\% & 7.4\% & 1.0\%     & 56.3\% & 69.1\% & \bf 65.7\% \\
        \citet{long2020pragmatic} &\CIRCLE&\Circle&\CIRCLE&\CIRCLE &0.0\%&0.0\%& -- & 2.2\% & 4.7\% &  --  & 53.5\% & 54.5\% & -- \\
        \citet{watson2021importance} &\CIRCLE&\Circle&\CIRCLE&\CIRCLE & 0.1\% & 0.9\% & 0.02\% & 1.3\% & 5.4\% & 1.1\% & 59.1\% & 70.1\% & 65.4\% \\
        {\citet{ye2021enhanced}}
        &\CIRCLE&\Circle&\CIRCLE&\CIRCLE & - & - & - & - & - & - & {60.3\%} & {76.9\%} & {65.5\%} \\
        \midrule
        Ours &\CIRCLE&\CIRCLE&\CIRCLE&\CIRCLE &\textbf{2.2\%}&\textbf{11.2\%}& \textbf{0.09\%} & \bf 8.4\% &  \bf 27.6\% & \bf 1.4\% &  \bf 63.8\% &  \bf 82.6\% & 65.6\% \\
        \bottomrule
    \end{tabular}
    \vspace{.5em}
    \caption{\textbf{Comparison of prior membership inference attacks} under the same settings
    for well-generalizing models on CIFAR-10, CIFAR-100, and WikiText-103 {using 256 shadow models}.
    Accuracy is only presented for completeness; we do not believe this is
    a meaningful metric for evaluating membership inference attacks.
    Full ROC curves are presented in Appendix~\ref{sec:appendix}.
    }
    \label{tab:prior_work}
\end{table*}

\subsection{Offline attack evaluation}
\label{ssec:no_out}

\Cref{fig:noin} evaluates our offline attack from \Cref{ssec:estimating_likelihood}, where the adversary performs the costly operations of training shadow models only before being handed the target query point $(x,y)$. 
%
%
Our attack performs only slightly worse in this offline setting---at an FPR of $0.1\%$, our offline attack's TPR is at most $20\%$ lower than that of our best online attack with the same number of shadow models.

\subsection{Re-evaluating prior membership inference attacks}
%
In order to understand how our attack compares to prior work, 
we now re-evaluate prior attack techniques under our low-FPR objective.
We study these attacks following the same evaluation protocol
introduced above and on the same datasets (for WikiText, we omit a few
entries for attacks that are not directly applicable to sequential language models).

A summary of our analysis is presented in \Cref{tab:prior_work}.
We compare the efficacy of eight representative attacks from the literature. For each attack, we compute a full ROC curve and select a decision threshold that maximizes TPR at a given FPR. %
%
Surprisingly, we find that despite being published in 2019, the 
attack of \citet{sablayrolles2019white} outperforms
other attacks under our metric (often by an order of magnitude), even when compared to more
recent attacks such as \citet{jayaraman2020revisiting} (PETS'21) and \citet{song2021systematic} (USENIX'21).

\bigskip \noindent
\textbf{Shadow models.}
One of the first membership inference attacks (due to \citet{shokri2016membership})
that improves on the baseline
LOSS attack,
introduced the idea of shadow models,
but used in a simpler way than we have done here.
Each shadow model $f_i$ (of a similar type to the target model $f$) is trained on random subsets $D_i$ of training data available to the adversary. 
The attack then trains a new neural network $g$ to predict an example's membership status. Given the pre-softmax features $f_i(x)$ and class label $y$, the model $g$ predicts whether the data point $(x, y)$ was a member of the shadow training set $D_i$. For a target model $f$ and point $(x, y)$, the attack then outputs $g(f(x), y)$ as a membership confidence score.

We implement this by training shadow models that randomly subsample half of the total dataset. The training set of the shadow models thus partially overlaps with the training set of the target model $f$. This is a stronger assumption than that made by \citet{shokri2016membership} and thus yields a slightly stronger attack.
Despite being significantly more expensive than the LOSS attack due to the
overhead of training many shadow models and then training a membership
inference predictor on the output of the models, this attack does 
not perform significantly better at low false-positive rates.

\bigskip \noindent
\textbf{Multiple queries.} 
It is possible to improve attacks by making multiple queries to the model.
\citet{jayaraman2020revisiting} do this with their MERLIN attack, that queries the target model $f$ multiple times on a sample $x$ perturbed with fresh Gaussian noise, and measures how the model's loss varies in the neighborhood of $x$.
However, even when querying the target model $100$ times and carefully choosing the noise magnitude, we find that this attack does not improve the adversary's success at low false-positive rates.

\citet{choquette2021label} suggest an alternate technique to increase attack accuracy
when models are trained with \emph{data augmentations}. In addition to querying the model on $f(x)$, 
this attack also queries on augmentations of $x$ that the model might have seen during training.
This is the direct motivation for us making these additional queries,
which as we will show in Section \ref{ssec:num_queries} improves our attack success rate considerably.

%

\bigskip \noindent 
\textbf{Per-class hardness.} 
Instead of using per-example hardness scores as we have done, a potentially simpler
method would be to 
design just one scoring function $\mathcal{A}'_y$ per class $y$, 
by scaling the model's loss by a class-dependent value: $\mathcal{A'}_y(x, y) = \mathcal{A'}(x, y) - \tau_y$.
For example, in the ImageNet dataset~\cite{deng2009imagenet} there are several hundred classes for
various breeds of dogs, and so correctly classifying individual dog breeds tends to be harder 
than other broader classes.
Interestingly, despite this intuition, in practice using
per-class thresholds neither helps improve balanced attack accuracy 
nor attack success rates at low false-positive rates, although it does
improve the AUC of attacks on CIFAR-10 and CIFAR-100 by $2\%$.

The attack of \citet{song2021systematic} reported in \Cref{fig:full_comparison} and \Cref{tab:prior_work}
combines per-class scores with additional techniques. Instead of working with the standard cross-entropy loss, this attack uses a \emph{modified entropy} measure and trains shadow models to approximate the distributions of entropy values for members and non-members of each class. Given a model $f$ and target sample $(x, y)$, the attack computes a hypothesis test between the (per-class) member and non-member distributions
(see~\cite{song2021systematic}). Despite these additional techniques, this attack does not improve upon the baseline attack \cite{shokri2016membership} at low FPRs.

\bigskip \noindent
\textbf{Per-example hardness.} 
As we do in our work,
a final direction considers per-example hardness.
\citet{sablayrolles2019white} is the most direct influence for LiRA. 
Their attack,
$\mathcal{A}'(x,y) = \ell(f(x), y) - \tau_{x, y}$, 
scales the loss by a per-example hardness threshold $\tau_{x, y}$ that is estimated by training shadow models. 
Instead of fitting Gaussians to the shadow models' outputs as we do, this paper takes a simpler non-parametric approach and
sets the threshold near the midpoint
$\tau_{x,y} = (\mu_{\text{in}}(x,y) + \mu_{\text{out}}(x,y))/2$
so as to maximize the attack accuracy;
here $\mu_{\text{in}},\mu_{\text{out}}$ are the means computed as we do.

The recent work of \citet{watson2021importance} considers an offline variant of \citet{sablayrolles2019white}, that sets $\tau_{x,y} = \mu_{\text{out}}(x,y)$ (i.e., each example's loss is calibrated by the average loss of shadow models not trained on this example).

Both \citeauthor{sablayrolles2019white} and \citeauthor{watson2021importance} evaluate their attacks using average case metrics (balanced accuracy and AUC),
and find that using per-example hardness thresholds can moderately improve upon past attacks. In our evaluation (\Cref{tab:prior_work}), we find that the balanced accuracy and AUC of their approaches are actually slightly \emph{lower} than those of other simpler attacks. Yet, we find that per-example hardness-calibrated attacks reach a \emph{significantly better} true-positive rate at low false-positive rates---and are thus much better attacks
according to our suggested evaluation methodology.

The discrepancy between the balanced accuracy and our recommended low false-positive metric is even more stark for
the attack of \citet{long2020pragmatic}. 
This attack also trains shadow models 
to estimate per-example hardness, but additionally filters out a fraction of outliers
to which the attack should be applied, and then makes no confident guesses for non-outliers. 
This attack thus
cannot achieve a high average accuracy, yet outperforms most prior attacks 
at low false-positive rates.

To expand, this attack \cite{long2020pragmatic} builds a graph of all examples $x$, where an edge between $x$ and $x'$ is weighted by the cosine similarity between the features $z(x)$ and $z(x')$. Our implementation of this attack selects the $10\%$ of outliers with the largest distance to their nearest neighbor in this graph.
For each such outlier $(x,y)$, the attack trains shadow models to numerically estimate the probability of observing a loss as high as $\ell(f(x), y)$ when $(x,y)$ is not a member.

{
The attack in the concurrent work of \citet{ye2021enhanced} is close in spirit to ours. They follow the same approach as our offline attack, by training multiple OUT models and then performing an exact one-sided hypothesis test. Specifically, to target an FPR of $\alpha$, their attack sets each example's decision threshold so that an $\alpha$-fraction of the measured OUT losses for that example lie below the threshold.
}

{
The critical difference between our attack and these prior attacks is that we use a more efficient \emph{parametric} approach, that models the distribution of losses as Gaussians. Since \citet{sablayrolles2019white} and \citet{watson2021importance} only measure the means of the distributions, the attacks are sub-optimal if different samples' loss distributions have very different scales and spreads (c.f. \Cref{fig:logits}).
The attacks of \citet{long2020pragmatic} and \citet{ye2021enhanced} take into account the full distribution of OUT losses, but have difficulties extrapolating to low FPRs due to the lack of a parametric assumption.
By design, the exact test of \citet{ye2021enhanced} can at best target an FPR of $1/N$ with $N$ shadow models. It is thus inapplicable in the setting we consider here (256 shadow models, and a target FPR of $0.1\%$).
\citet{long2020pragmatic} extrapolate to the tails of the empirical loss distribution using cubic splines, which easily overfit and diverge outside of their support.
}

\begin{figure}[t]
    \centering
    \includegraphics[scale=.7]{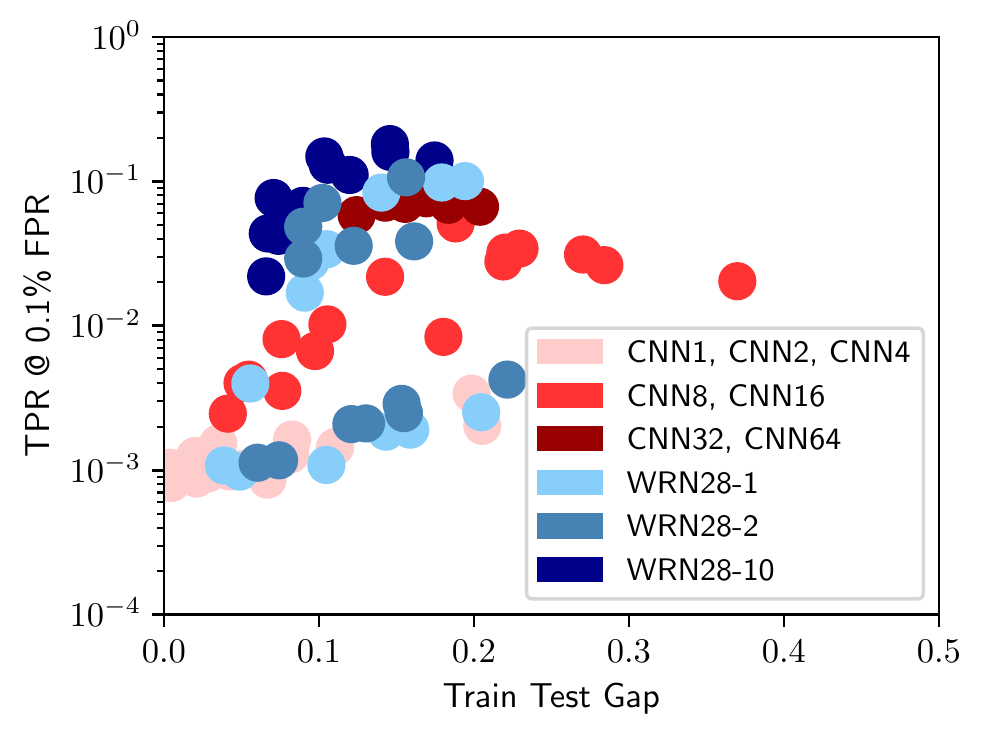}
    \caption{Attack true-positive rate versus model train-test gap for a variety of CIFAR-10 models.}
    \label{fig:tpr0p1_gap}
\end{figure}

\subsection{Membership inference and overfitting}
\label{ssec:overfitting}

To better understand the relationship between overfitting and vulnerability
to membership inference attacks, Figure~\ref{fig:tpr0p1_gap} plots
various models' train-test gap (that is, their train accuracy minus
their test accuracy) versus our attack's TPR at an FPR of $0.1\%$.
We train CNN models and Wide ResNets (WRN) of various sizes on CIFAR-10, with different optimizers and data augmentations (see \Cref{ssec:mismatched_training} for details). Each point represents one training configuration for the target model.

While there is an overall trend that overfit models (those with higher train-test gap) are more vulnerable to attack, we do find examples of models
that have \textbf{identical train-test gaps but are $\textbf{100}\times$ more vulnerable to attack}.
In \Cref{fig:tpr0p1_test} in the Appendix we further plot the attack TPR as a function of the test accuracy of these models. There, we observe a clear trend that \textbf{more accurate models are more vulnerable to attack}.

\begin{table}[t]
    \centering
    \begin{tabular}{@{}lr@{}}
    \toprule
        \textbf{Attack Approach} & \textbf{TPR @ 0.1\% FPR} \\
        \midrule
        LOSS attack~\cite{yeom2018privacy} & 0.0\% \\
        \quad + Logit scaling & 0.1\% \\
        \quad + Multiple queries & 0.1\% \\
        \midrule
        LOSS attack~\cite{yeom2018privacy} & 0.0\% \\
        \quad + Per-example thresholds ($\tilde{\mathbb{Q}}_{\text{out}}$ only)~\cite{watson2021importance} & 1.3\% \\
        \quad + Logit scaling & 4.7\% \\
        \quad + Gaussian Likelihood & 4.7\% \\
        \quad + Multiple queries \textbf{(our offline attack)} & \textbf{7.1\%} \\
        \midrule
        LOSS attack~\cite{yeom2018privacy} & 0.0\% \\
        \quad + Per-example thresholds ($\tilde{\mathbb{Q}}_{\text{in}}$ \& $\tilde{\mathbb{Q}}_{\text{out}}$)~\cite{sablayrolles2019white} & 1.7\% \\
        \quad + Logit scaling & 1.9\% \\
        \quad + Gaussian Likelihood & 5.6\% \\
        \quad + Multiple queries \textbf{(our online attack)} & \textbf{8.4\%} \\
    \bottomrule
    \end{tabular}
    \caption{By iteratively adding the main components of our
    attack we can interpolate between the simple LOSS threshold attack \cite{yeom2018privacy} and our full offline and online attacks. 
    }
    \label{tab:ablation_summary}
\end{table}

\section{Ablation Study}
\label{sec:ablation}

Our attack has a number of moving pieces that are connected in various
ways; in this section we investigate how these pieces
come together to reach such high accuracy at low false-positive rates.
We exclusively use CIFAR-10 for these ablation studies as it is
the most popular image classification dataset and is the
hardest datasets we have considered;

A summary of our analysis is presented in Table~\ref{tab:ablation_summary}.
The baseline LOSS attack achieves a true-positive rate of $0\%$ at a false
positive rate of $0.1\%$ (as shown previously in \Cref{fig:roc_curve_log}).
If we do not use per-example thresholds, this basic attack can only be marginally
improved by properly scaling the loss and issuing multiple queries to the target model.

By incorporating per-example thresholds obtained by estimating the distributions $\tilde{\mathbb{Q}}_{\text{in}}$ and $\tilde{\mathbb{Q}}_{\text{out}}$ as in~\cite{sablayrolles2019white}, the attack success rate increases to $1.7\%$---about one-order-of-magnitude better than chance.
By ensuring that we appropriately re-scale the model losses
(explored in detail in \Cref{ssec:loss_function}) and fitting the re-scaled losses with Gaussians (see \Cref{ssec:global_variance}), we increase the attack success rate
by a factor of $3.3\times$.
Finally, we can nearly double the attack success rate by evaluating the target model
on the same data augmentations as used during training, as we will show in
\Cref{ssec:num_queries}.

We also perform the same ablation but with the offline variant of our attack.
Here, if we start with the attack of \citet{watson2021importance} 
to reach a $1.3\%$ true-positive rate;
adding logit scaling, Gaussian likelihood, and multiple queries yields an attack that is nearly as strong as our full attack (TPR of $7.1\%$ versus $8.4\%$ at an FPR of $0.1\%$).

\subsection{Logit scaling the loss function}
\label{ssec:loss_function}

\begin{figure}
    \centering
    \includegraphics[scale=.75]{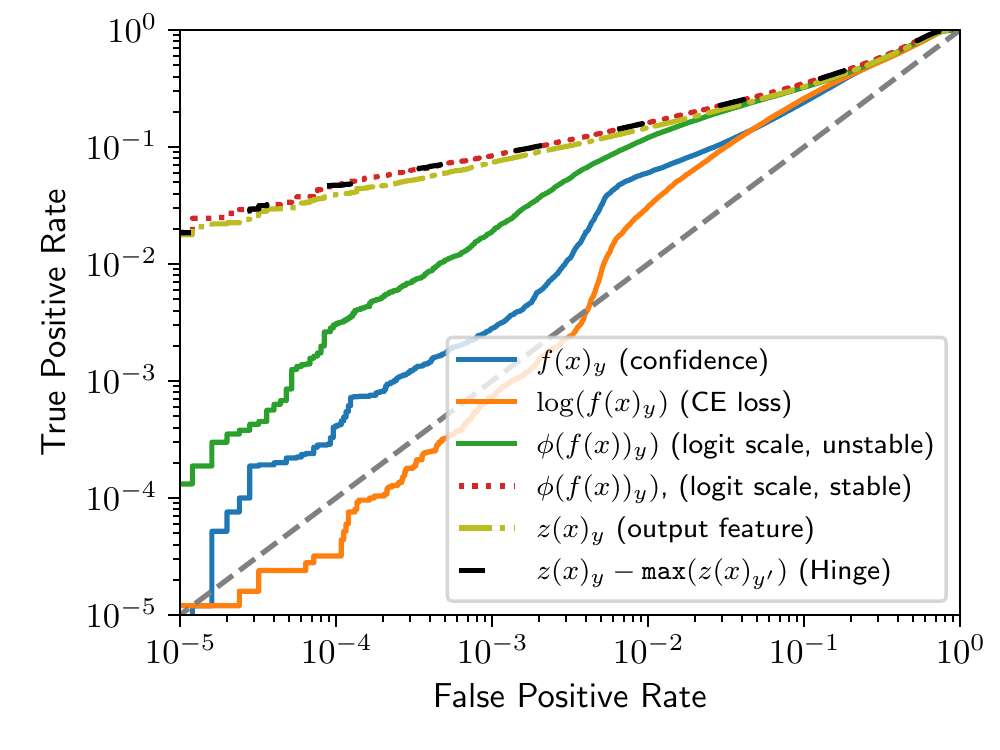}
    \caption{The best scoring metrics ensure the output distribution is approximately
    Gaussian, and the worst metrics are not easily modeled with a
    standard distribution (see Figure~\ref{fig:logits}).}
    \label{fig:numerics}
\end{figure}

The first step of our attack projects the model's confidences to a logit scale to ensure that the distributions that we work with are approximately normal. 
%
\Cref{fig:numerics} compares performance of our attack for various choices of statistics that we can fit using shadow models.
Recall that we defined our neural network function $f(x)$ to denote the evaluation of the
model along with a final softmax activation function;
we use $z(x)$ to denote the pre-softmax activations of the neural network.

As expected, we find that using the model's confidence $f(x)_y \in [0,1]$, or its logarithm (the cross-entropy loss), leads to poor performance of the attack since these statistics do not behave like Gaussians (recall from \Cref{fig:logits}).

Our logit rescaling performs best, but the exact numerical computation of the logit function $\phi(p)=\log(\frac{p}{1-p})$ matters. We consider two mathematically equivalent variants:
\begin{align*}
    \phi_{\text{unstable}} &= \log(f(x)_y) - \log(1-f(x)_y)\\
    \phi_{\text{stable}} &= \log(f(x)_y) - \log\sum_{y' \neq y} f(x)_{y'} \;.
\end{align*}
We find that the second version is more stable in practice, when the model's confidence is very high, $f(x)_y \approx 1$ (we compute all logarithms as $\log(x+\epsilon)$ for a small $\epsilon>0$). Note that this second stable variant requires access to the full vector of model confidences $f(x) \in [0,1]^n$ rather than just the confidence of the predicted class.

If the adversary can query the model to obtain the unnormalized features $z(x)$ (i.e., the outputs of the model's last layer before the softmax function),  a hinge loss performs similarly
\[
\ell_{\text{Hinge}}(x, y) = z(x)_y - \max_{y' \neq y} z(x)_{y'} \;.
\]
To see why this is the case, observe that
\begin{align*}
    \phi(f(x)_y) &= \log(f(x)_y) - \log\sum_{y' \neq y} f(x)_{y'}\\
    &= z(x)_y - \underset{y' \neq y}{\operatorname{LogSumExp}}\ z(x)_{y'} \;,
\end{align*}
where the $\operatorname{LogSumExp}$ function is a smooth approximation to the maximum function. When the features $z(x)$ are available, \textbf{we recommend using the hinge loss} as its computation is numerically simpler than that of the logit-scaled confidence.

We note that the different attack variants we consider here lead to orders-of-magnitude differences in attack performance at low false-positive rates---even though all variants achieve similar AUC scores (68--72\%).
This again highlights the importance of carefully designing attacks,
and of measuring attack performance at low false-positive rates rather than on average across the entire ROC curve.

The choice of an appropriate loss function can also have a major impact on previous MIAs. For example, for the attack of \citet{watson2021importance} (which scales the model's loss by the mean loss of OUT models not trained on the example, $\mu_{\text{out}}(x,y)$) applying logit scaling nearly quadruples the attack's true-positive rate at an FPR of 0.1\% (see Table~\ref{tab:ablation_summary}).

\subsection{Gaussian distribution fitting}
\label{ssec:global_variance}

\begin{figure}[t]
    \centering
    \includegraphics[scale=.75]{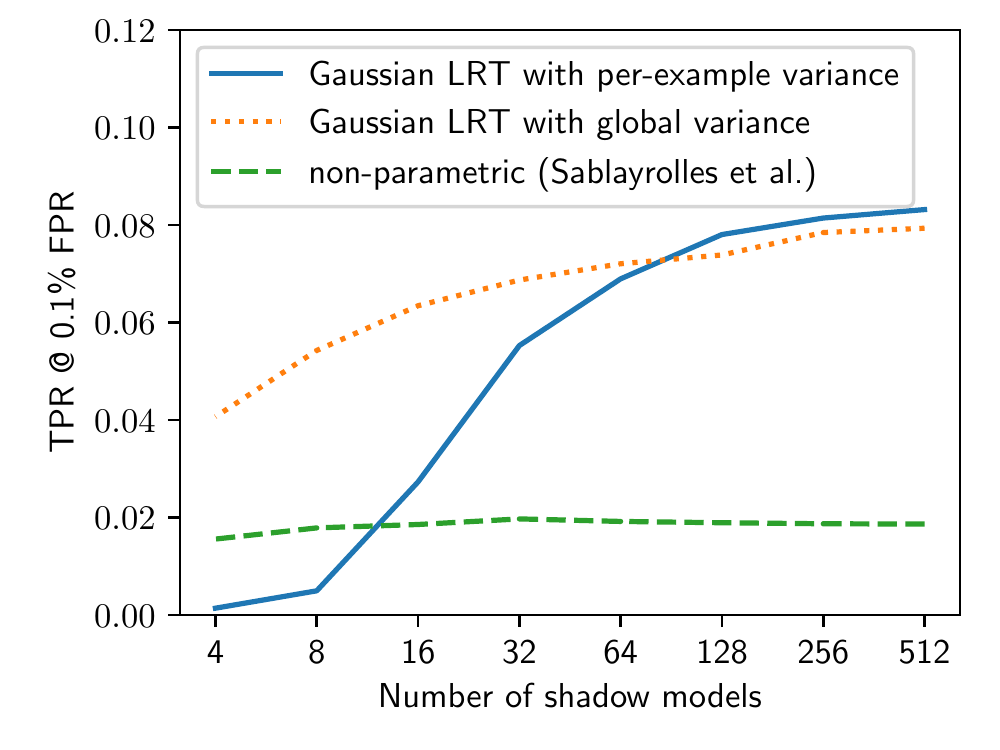}
    \caption{Attack success rate increases as the number of shadow models increases, with the benefit eventually tapering off.
    When fewer than 64 models are used, it is better to estimate the variance of the model confidence as a global parameter instead of computing it on a per-example basis.}
    \label{fig:model_num}
\end{figure}

Like other shadow models membership inference attacks~\cite{shokri2016membership},
our attack requires that we train enough models to accurately estimate the distribution of losses.
It is thus desirable to minimize the number of shadow models that are necessary.
However, most prior works in \Cref{tab:prior_work} that rely on shadow models do not analyze this tradeoff and report results only for a fixed number of shadow models~\cite{shokri2016membership, long2020pragmatic, sablayrolles2019white}. 

\Cref{fig:model_num} displays our online attack's TPR at a fixed FPR of 0.1\%, as we vary the number of shadow models (half IN and half OUT). 
Training more than $64$ shadow models provides diminishing benefits, but the attack deteriorates quickly with fewer models---due to the difficulty of fitting Gaussian distributions on a small number of data points.

With a small number of shadow models, we can improve the attack considerably by estimating the variances $\sigma_{\text{in}}^2$ and $\sigma_{\text{out}}^2$ of model confidences in \Cref{alg:main} \emph{globally} rather than for each individual example. 
That is, we still estimate the means $\mu_{\text{in}}$ and $\mu_{\text{out}}$ separately for each example, but we estimate the variance $\sigma_{\text{in}}^2$ (respectively $\sigma_{\text{out}}^2$) over the shadow models' confidences on \emph{all} training set members (respectively non-members).

For a small number of shadow models ($<64$), estimating a global variance outperforms our general attack that estimates the variance for each example separately. For a larger number of models, our full attack is stronger: with $1024$ shadow models for example, the TPR decreases from
$8.4\%$ to $7.9\%$ by using a global variance.



\subsection{Number of queries}
\label{ssec:num_queries}

Models are typically trained to minimize their loss not only on the original training
example, but also on \emph{augmented} versions of the example.
It therefore makes sense to perform membership inference attacks on the augmented
versions of the example that may have been seen during training.
Results of this analysis are presented in Table~\ref{fig:numaugmentations}.
There are $162$ potential augmentations of each training
image for our CIFAR-10 model ($2 \times 9 \times 9$, computed by either horizontally flipping the image or not,
and shifting the image by up to $\pm4$ pixels in each height or width). We find that 
querying on just $2$ augmentations gives most of the benefit, with
increasing to 18 queries performing identically to all $162$ augmentations.

\begin{table}[]
    \centering
    \begin{tabular}{@{}lrr@{}}
    \toprule
         & \multicolumn{2}{c}{\textbf{TPR @ FPR}} \\
        \cmidrule(l{5pt}){2-3}
        \textbf{Queries} & 0.1\% & 0.001\% \\
        \midrule
        1 (no augmentations) & 5.6\% & 1.0\% \\
        2 (mirror) & 7.5\% & 1.8\% \\
        18 (mirror + shifts) & \textbf{8.4\%} & \textbf{2.2\%} \\
        162 (mirror + shifts) & \textbf{8.4\%} & \textbf{2.2\%} \\
        \bottomrule
    \end{tabular}
    \caption{Querying on augmented versions of the image doubles the true-positive rate
    at low false-positive rates, with most benefits given by just two queries.}
    \label{fig:numaugmentations}
\end{table}

\subsection{Disjoint datasets}
\label{ssec:mismatched_data}

In our experiments so far, we trained both the target models and the adversary's shadow models by subsampling from a common dataset. That is, we use a large dataset $D_{\text{attack}}$ (e.g., the entire CIFAR-10 dataset) to train shadow models, and the target model's training set $D_{\text{train}}$ is some (unknown) subset of this dataset.
This setup favors the attacker, as the training sets of shadow models and the target model can partially overlap. 

In a real attack, the adversary likely has access to a dataset $D_{\text{attack}}$ that is \emph{disjoint} from the training set $D_{\text{train}}$. We now show that this more realistic setup has only a minor influence on the attack's success rate.

For this experiment, we use the CINIC-10 dataset~\cite{darlow2018cinic}. This dataset combines CIFAR-10 with an additional 210k images taken from ImageNet that correspond to classes contained in CIFAR-10 (e.g., bird/airplane/truck etc). We train a target model and 128 shadow models (OUT models only) each on 50{,}000 points. We compare three attack setups:
\begin{enumerate}
    \item The shadow models' training sets are sampled from the full CINIC-10 dataset. This is the same setup as in all our previous experiments, where $D_{\text{train}} \subset D_{\text{attack}}$.
    \item The shadow models' training sets have no overlap with the target model, i.e., $D_{\text{train}} \cap D_{\text{attack}} = \emptyset$.
    \item The target model is trained on CIFAR-10, while the attacker trains shadow models on the ImageNet portion of CINIC-10. There is thus a \emph{distribution shift} between the target model's dataset and the attacker's dataset. 
\end{enumerate}

\begin{figure}
    \centering
    \includegraphics[scale=.75]{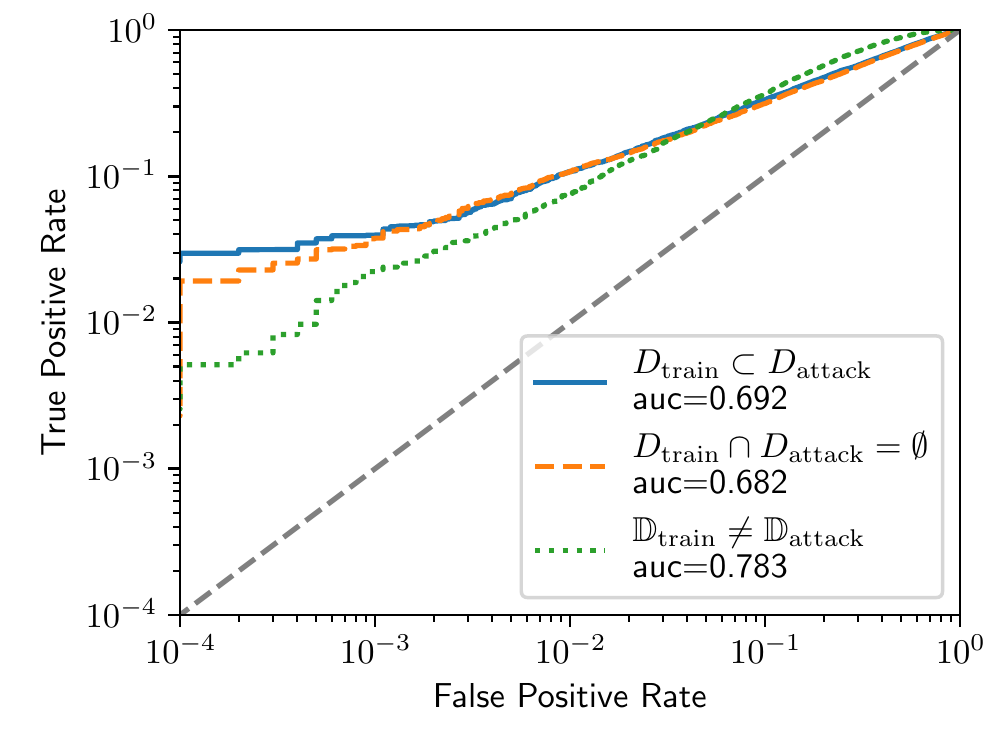}
    \caption{The attack's success rate on CINIC-10 remains unchanged when the training sets of shadow models are sampled from a dataset $D_{\text{attack}}$ that is disjoint from the target model's training set $D_{\text{train}}$. The attack's performance does decrease when the two datasets are sampled from different \emph{distributions}.}
    \label{fig:cinic_dataset}
\end{figure}

\Cref{fig:cinic_dataset} shows that our attack's performance is not influenced by an overlap between the training sets of the target model and shadow models. 
The attack success is unchanged when the attacker uses a disjoint dataset.
A \emph{distribution shift} between the training sets of the target model and shadow models does reduce the attack's TPR. 
Surprisingly, the attack's AUC is much higher when there is a distribution shift---we leave an explanation of this phenomenon to future work.                                                                                                                                                                 

\begin{figure*}[t]
    \centering
    \begin{subfigure}[t]{0.4\textwidth}
    \centering
    \includegraphics[height=1.5in]{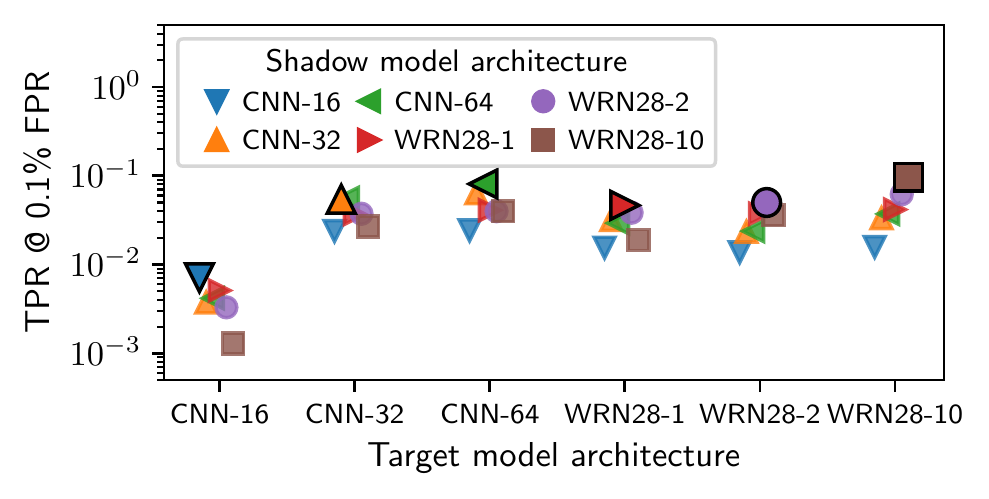}
    \caption{Vary model architecture.}
    \label{fig:mismatch_archi}
    \end{subfigure}
    \hfill
    \begin{subfigure}[t]{0.23\textwidth}
    \centering
    \includegraphics[height=1.5in]{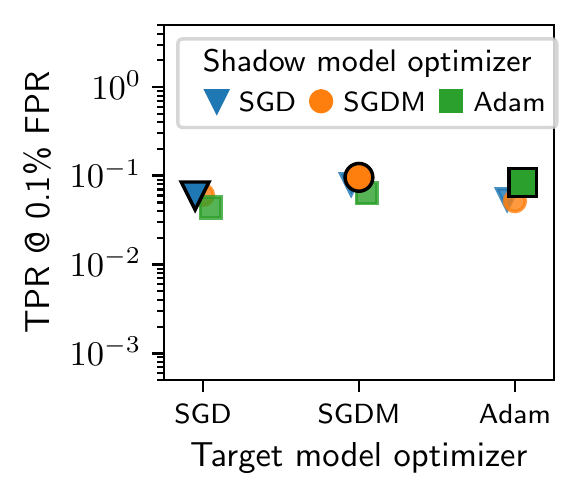}
    \caption{Vary training optimizer.}
    \label{fig:mismatch_opt}
    \end{subfigure}
    \hfill
    \begin{subfigure}[t]{0.33\textwidth}
    \centering
    \includegraphics[height=1.5in]{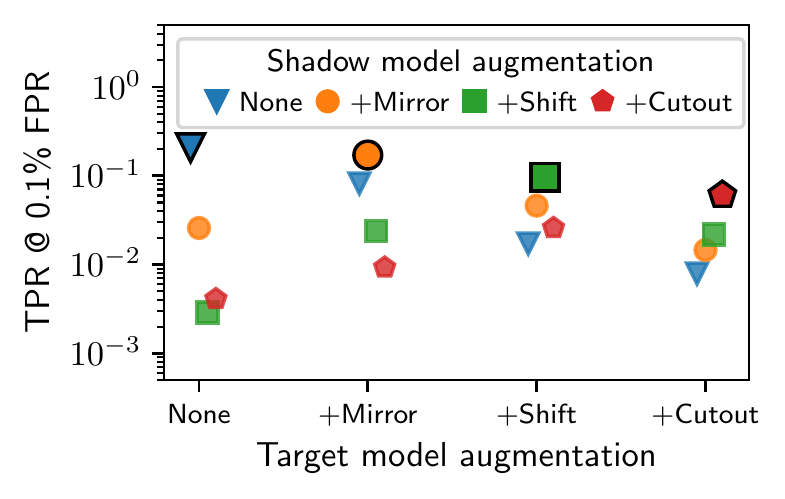}
    \caption{Vary data augmentation.}
    \label{fig:mismatch_aug}
    \end{subfigure}
    \caption{Our attack succeeds when the adversary is uncertain of the target model's training setup. We vary the target model's architecture (a), the training optimizer (b) and the data augmentation (c), as well as the adversary's guess of each of these properties when training shadow models. The attack performs best when the adversary guesses correctly (black-lined markers).
    }
    \label{fig:mismatch}
\end{figure*}

\subsection{Mismatched training procedures}
\label{ssec:mismatched_training}

%
We now explore how our attack is affected if the attacker does not know the exact training procedure of the target model. 
We train models with various architectures, optimizers, and data augmentations to investigate the attack's performance when the adversary guesses each of these incorrectly. For each attack, we train $64$ shadow models and use our online attack variant with a global estimate of the variance (see \Cref{ssec:global_variance}).
\Cref{fig:mismatch} summarizes our results at a fixed FPR of $0.1\%$. 
Appendix \Cref{fig:architecture,fig:wrn28_optimizer,fig:wrn28_augmentation} have full ROC curves.

In \Cref{fig:mismatch_archi}, we vary the target model's architecture. We study three CNN models (with 16, 32 and 64 convolutional filters), and three Wide ResNets (WRN) with width 1, 2 and 10. All models are trained with SGD with momentum and with random augmentations. Our attack performs best when the attacker trains shadow models of the same architecture as the target model, but using a similar model (e.g., a WRN28-1 instead of a WRN28-2) has a minimal effect on the attack.
Moreover, we find that for both the CNN and WRN model families, \emph{larger models are more vulnerable to attacks}.

In \Cref{fig:mismatch_opt} we fix the architecture to a WRN28-10, and vary the training optimizer: SGD, SGDM (SGD with momentum) or Adam. For both the defender or the attacker, the choice of optimizer has minimal impact on the attack.

Finally, in \Cref{fig:mismatch_aug} we fix the architecture (WRN28-10) and optimizer (SGDM) and vary the data augmentation used for training: none, mirroring, mirroring + shifts, mirroring + shifts + cutout. The attacker's guess of the data augmentation is used both to train shadow models, and to create additional queries for the attack. 
We find that correctly guessing the target model's data augmentation has the highest impact on attack performance. Models trained with stronger augmentations are harder to attack, as these models are less overfit.

\begin{figure}
    \centering
    \vspace{-1em}
    \includegraphics[width=\columnwidth]{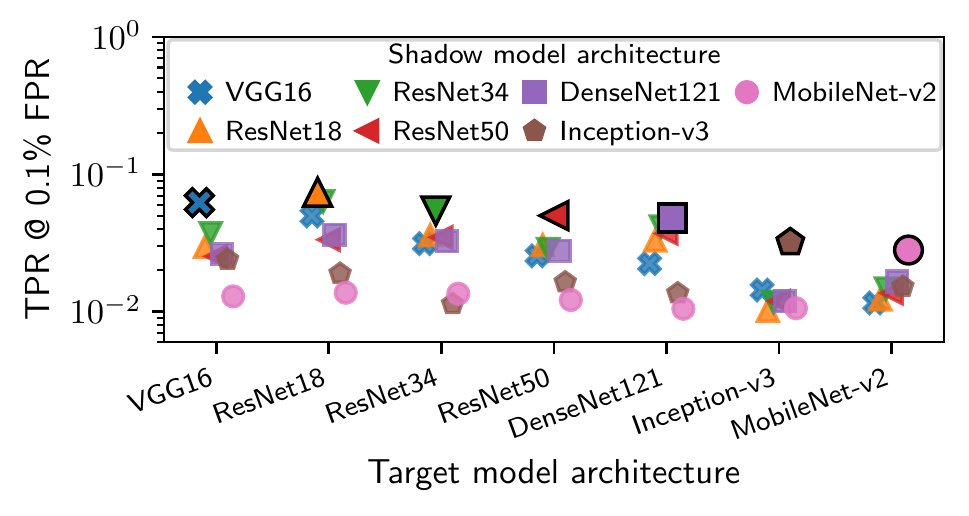}
    \caption{Our attack succeeds against real state-of-the-art CIFAR-10 models \cite{huy_phan_2021_4431043}. The attacker trains shadow models on a random subset of $50{,}000$ points from the entire CIFAR-10 dataset. The attack performs best when the shadow models have the same architecture as the target model, but training different models still leads to a strong attack.}
    \label{fig:real_world}
\end{figure}

\section{Additional Investigations}

We now pivot from evaluating our attack to using
our attack as a tool to better understand
memorization in real models (\S\ref{ssec:real_world})
and why memorization occurs (\S\ref{ssec:why}).

\subsection{Attacking real-world models}
\label{ssec:real_world}

All our experiments so far have involved attacking models that we ourselves have trained.
To ensure that we did not somehow train \emph{weakly accidentally private (or non-private)} models, we now show that our attacks also succeed on existing pre-trained state-of-the-art models. 
To this end, we load standard models pre-trained by \citet{huy_phan_2021_4431043} on the complete CIFAR-10 training set (50{,}000 examples). We train 256 shadow models by using the same training code and subsampling $50{,}000$ points at random from the entire CIFAR-10 dataset ($60{,}000$ examples). On average, we have $213$ IN models and $43$ OUT models per example. \Cref{fig:real_world} shows our attack's true-positive rate at a 0.1\% FPR for various canonical model architectures. 

We consider two attack variants: (1) the adversary knows the target model's architecture and uses it to train the shadow models; (2) the shadow models use a different architecture than the target model. Since we only have $43$ models to estimate the distribution $\tilde{\mathbb{Q}}_{\text{out}}$), esitmating a global variance for all examples performs best.
The results of this experiment are qualitatively similar to those in \Cref{ssec:mismatched_training}: (1) the model architecture has a small effect on the privacy leakage (e.g., the attack works better against a ResNet-18 than against a MobileNet-v2); (2) the attack works best when the shadow models share the same architecture as the target model, but it is robust to architecture mismatches. For example, attacking a ResNet-34 model with either ResNet-18 or ResNet-50 shadow models  leads to a minor drop in attack success rate (from 5\% TPR to 4\% TPR).

\begin{figure}[t]
    \centering
    \vspace{-2em}
    \includegraphics[scale=.75]{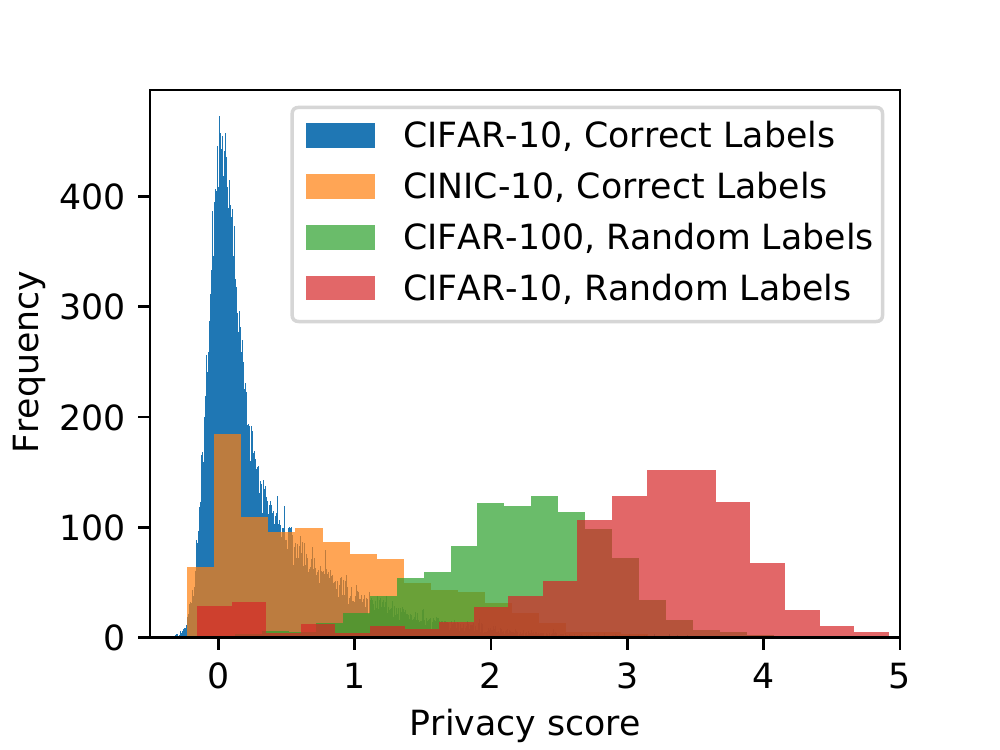}
    \caption{Out-of-distribution training examples are less private.}
    \label{fig:ood}
\end{figure}

\subsection{Why are some examples less private?}
\label{ssec:why}

While our average attack success rate is modest, the success rate at low false-positive rates can be very high.
This suggests that there is a subset of examples that are easier to
attack than others.
%
%
While a full investigation of this is beyond the scope of our paper,
we find an important factor behind why some samples
are less private is that they are out-of-distribution.

To make this argument,
we intentionally inject out-of-distribution examples into a model's training
dataset and compare the difficulty of attacking these newly inserted samples
versus typical examples.
Specifically, we insert $1{,}000$ examples from various out-of-distribution sources
into the $50{,}000$-example CIFAR-10 training dataset to form a new augmented
$51{,}000$ example dataset.
We then train shadow models on this dataset, run our attack,
and measure the distinguishability of distributions of losses for IN and OUT models for each of the $1{,}000$ newly inserted examples (we use a simple measure of distance between distributions here, defined as $d=\frac{|\mu_{\text{in}}-\mu_{\text{out}}|}{\sigma_{\text{in}} + \sigma_{\text{out}}}$).
\Cref{fig:ood} plots the distribution of these ``privacy scores'' assigned to each example.
As a baseline, in blue, we show the distribution of privacy 
scores for the standard CIFAR-10 dataset;
these are tightly concentrated around 0.

Next we show the privacy scores of examples inserted from the CINIC-10 dataset, which are drawn from ImageNet.
%
Due to this slight distribution shift, the CINIC-10 images have a larger privacy score on average: it is easier to detect their presence in the dataset because they are slightly out-of-distribution.

We can extend this further by inserting intentionally mislabeled images that are extremely out-of-distribution.
If we choose $1{,}000$ images (shown in red) from the CIFAR-10 test set and assign new random labels to each image, then we get a much higher privacy score for these images. 
Finally, we interpolate between the extreme OOD setting of random (and thus incorrectly) labeled CIFAR-10 images and correctly-labeled CINIC-10 by inserting randomly labeled images from CIFAR-100 (shown in green).
Because these images come from a disjoint class distribution, models will not typically be confident on their label one way or another unless they are seen during training.
The privacy scores here fall in between correctly labeled CINIC-10 and incorrectly labeled CIFAR-10.

\section{Conclusion}

As we have argued throughout this paper, membership inference attacks should focus
on the problem of achieving high true-positive rates at low false-positive rates.
Our attack presents one way to succeed at this goal.
There are a number of different evaluation directions that we hope future work will
explore under this direction.

\smallskip \noindent
\textbf{Membership inference attacks as a privacy metric.}
Both researchers \cite{murakonda2020ml} and practitioners \cite{song2020mia}
use membership inference attacks to measure privacy of trained models.
We argue that these metrics should use strong attacks (such as ours)
in order to accurately measure privacy leakage.
%
%
Future work using membership inference attacks should
consider the low false-positive rate regime,
to better understand if the privacy of even just a few users can be confidently breached.

\smallskip \noindent
\textbf{Usability improvements to membership inference attacks.}
The key limitation of per-example membership inference attacks is that they require
new hyperparameters that need to be learned from the data.
While it is much more important that attacks are
strong (even if slow) as opposed to fast (but weak),
we hope that future work will improve the computational efficiency
of our attack approach, in order to allow it to be deployed in more settings.

\smallskip \noindent
\textbf{Improving other privacy attacks with our method.}
Membership inference attacks form the basis for many other
privacy attack methods \cite{ganju2018property,carlini2019secret,carlini2020extracting}.
Our membership inference method, in principle, should be able to directly improve these attacks.

\smallskip \noindent
\textbf{Rethinking our current understanding of MIA results.}
The literature on membership inference attacks
has answered a number of memorization questions.
However, many (or even most) of these prior papers focused on the inadequate metric of
average-case attack success rates, instead of on the low false-positive rate regime.
As a result it will be necessary to re-investigate prior results from this perspective:
\begin{itemize}
    \item Do previously-``broken''  \cite{nasr2018machine,jia2019memguard}
    defenses prevent our attack?
    Prior defenses were only ever shown to be ineffective at
    preventing an adversary from succeeding on average---not confidently at low
    false-positive rates.
    
    \item How does differential privacy interact with our improved attacks? We have preliminary evidence that vacuous guarantees might prevent our low-FPR attacks (\Cref{ssec:dpsgd}). 
    
    \item Are attacks with reduced capabilities possible?
    For example, label-only attacks \cite{choquette2021label,rahimian2020sampling,li2020membership} can match the balanced accuracy of shadow-model approaches.
    But do these attacks work at low false-positive rates?
    
    \item Are attacks with extra capabilities more effective?
    Prior work has shown that access to gradient queries \cite{nasr2019comprehensive}
    or intermediate models \cite{salem2020updates} improves attack AUC.
    However, does this observation hold at low false-positive rates?
    
\end{itemize}

We hope that future work will be able to answer these questions, among many more,
in order to better evaluate (and develop) techniques that preserve the privacy of training data.
By developing attacks that succeed low false-positive rates,
we can evaluate privacy not as a measurement of the average user,
but of the most vulnerable.

\section*{Acknowledgements}
We are grateful to
Thomas Steinke, 
Dave Evans,
Reza Shokri,
Sanghyun Hong,
Alex Sablayrolles, 
Liwei Song,
Matthias L\'ecuyer
and the anonymous reviewers
for comments on drafts of this paper.

\bibliographystyle{plainnat}
{\small
  \bibliography{IEEEabrv,main}
  }

\appendices
\section{Additional Experiments}
\label{sec:appendix}

\subsection{Attacking DP-SGD}
\label{ssec:dpsgd}

Machine learning with differential privacy~\cite{abadi2016deep} is the main defence mechanism against privacy attacks including membership inference against machine learning models. Differential privacy provides an upper bound on the success of any membership inference attack. Recent works~\cite{nasr2021adversary,jagielski2020auditing} thus used membership attacks to empirically audit differential privacy bounds, in particular those obtained from DP-SGD~\cite{abadi2016deep}. In this work, we are interested in the effect of DP-SGD on the performance of our membership inference attack.

\begin{figure}[btp]
\centering

\begin{subfigure}[b]{0.32\textwidth}
\centering
\includegraphics[width=\textwidth]{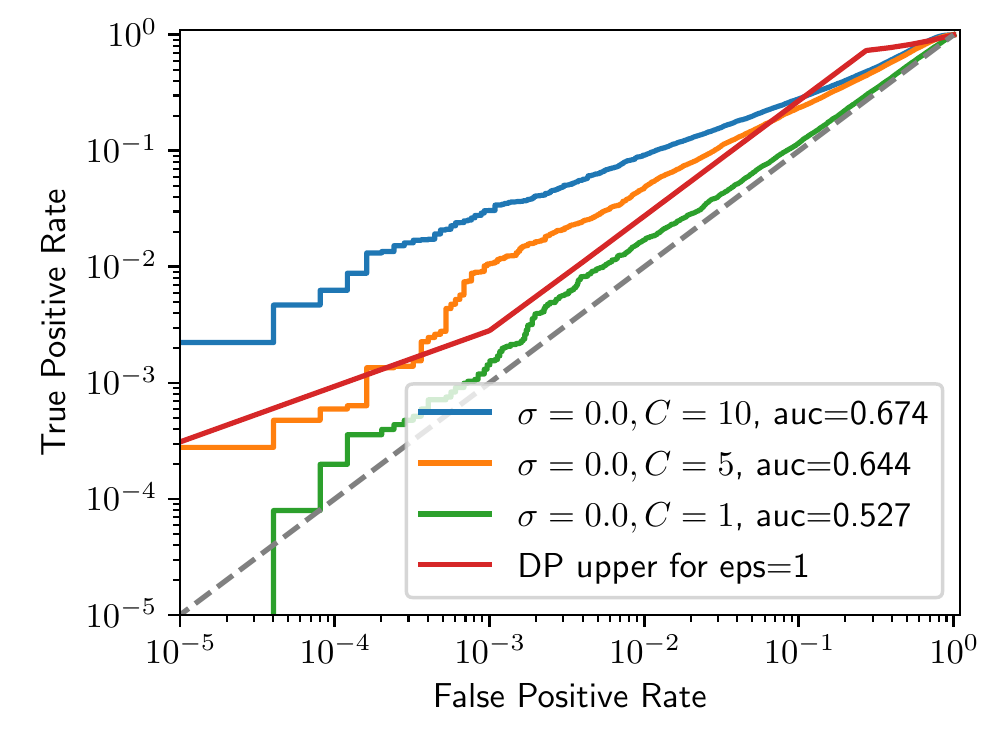}
\caption{$\epsilon = \infty$}
\end{subfigure}
\begin{subfigure}[b]{0.32\textwidth}
\centering
\includegraphics[width=\textwidth]{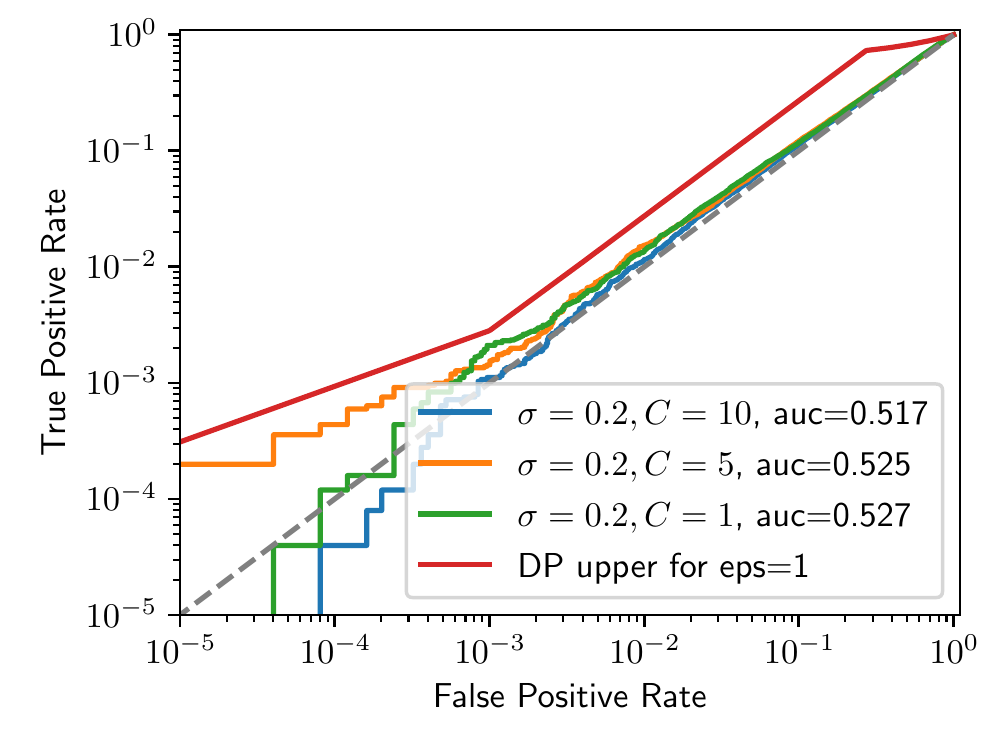}
\caption{$\epsilon > 5000$}
\end{subfigure}
\begin{subfigure}[b]{0.32\textwidth}
\centering
\includegraphics[width=\textwidth]{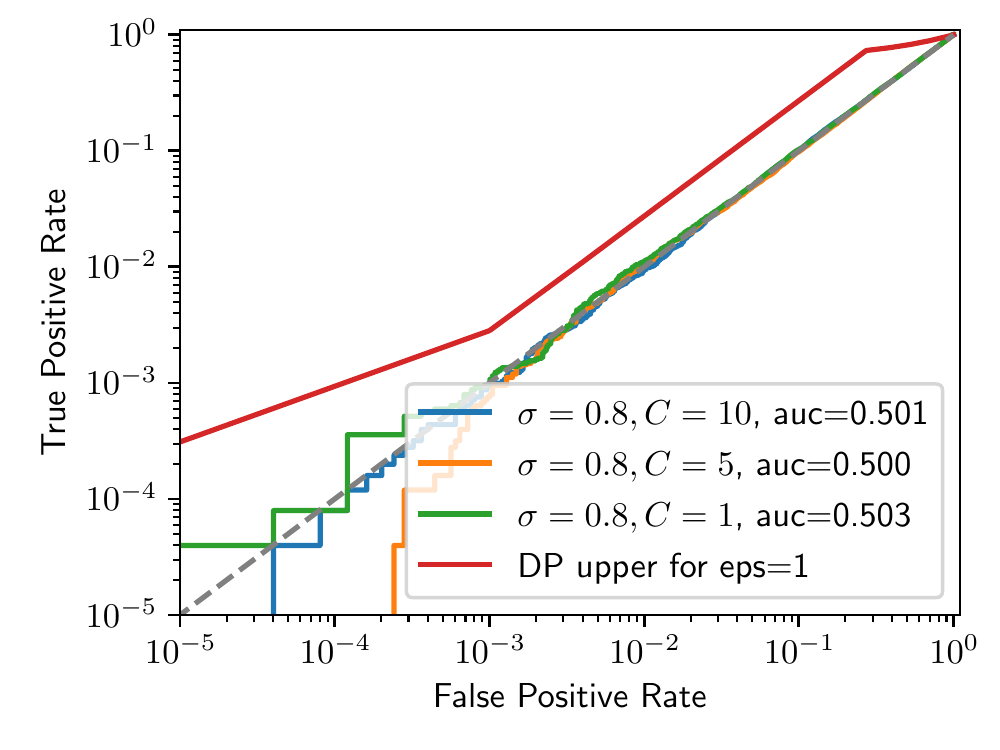}
\caption{$\epsilon = 8$}
\end{subfigure}
\caption{Effectiveness of using DP-SGD against our attack with different privacy budgets.}
\label{fig:dpsgd}
\end{figure}

We consider different combinations of DP-SGD's noise multiplier and clipping norm parameters in our evaluation. Table~\ref{tab:dpsgd_acc} summarizes the average accuracy of standard CNN models trained on CIFAR-10 with DP-SGD for different parameter sets. We evaluate the effectiveness of our membership inference attacks for these settings in Figure~\ref{fig:dpsgd}. Even just clipping the gradient norm without adding any noise reduces the performance of our attack significantly. However, small clipping norms can reduce the accuracy of the models as shown in Table~\ref{tab:dpsgd_acc}. 

\begin{table}[h]
    \centering
    \caption{Accuracy of the models trained with DP-SGD on CIFAR10 with different noise parameters}
    \begin{tabular}{@{}llll@{}}
    \toprule
        Noise Multiplier ($\sigma$) & $C = 10$ & $C=5$  & $C=1$  \\
        \midrule
        0.0 &  84.0\% & 78.5\%  & 61.3\%\\ 
        0.2 & 73.9\%  & 77.1\% & 62.8\%  \\
        0.8 & 36.9\% & 43.3\%  & 61.3\%  \\
        \bottomrule
    \end{tabular}
    
    \label{tab:dpsgd_acc}
\end{table}

For higher clipping norms, adding very small amounts of noise (Figure~\ref{fig:dpsgd}-b) reduces the effectiveness of the membership inference attack to chance, while resulting in models with higher accuracy.

Training models with very small amounts of noise is an effective defense against our membership inference attack, despite resulting in very large provable DP bounds $\epsilon$.

\clearpage
\subsection{White-box Attacks}
\begin{figure}[h]
    \centering
    \includegraphics[width=0.75\linewidth]{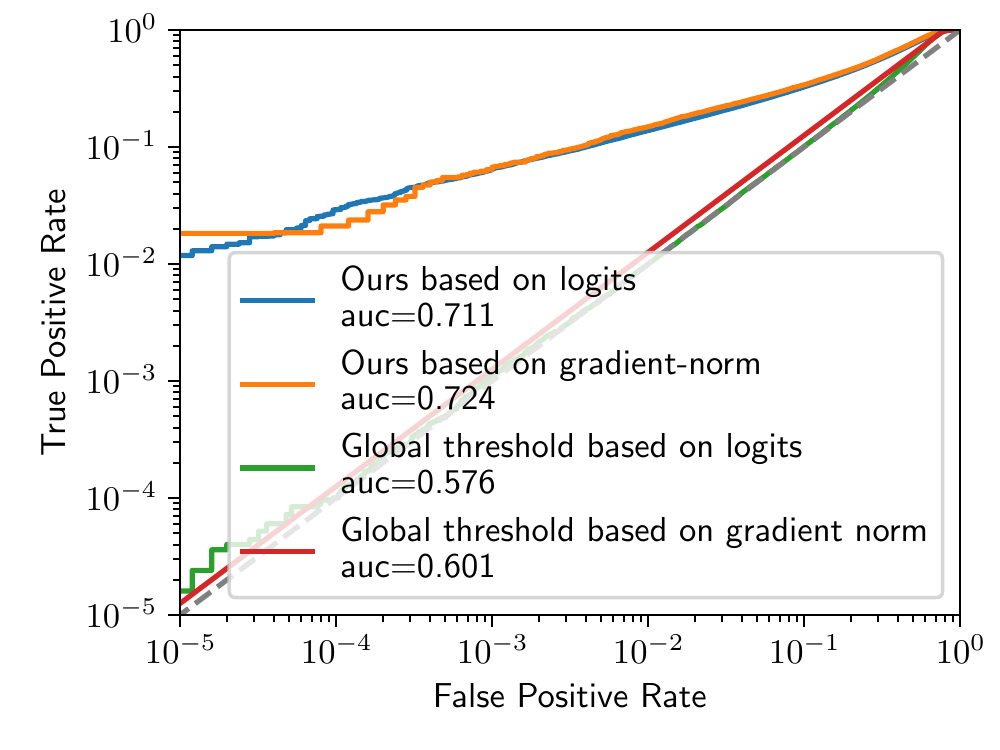}
    \caption{Comparison of the white-box attack using our approach to the black-box setting.
    }
    \label{fig:whitebox_log}
\end{figure}

Previous works~\cite{nasr2019comprehensive,song2019privacy} suggested that is possible to achieve better membership inference if the adversary has white-box access to the target model. In particular, previous works showed that using the norm of the model's gradient at a target point could increase the balanced accuracy of membership inference attacks. Figure~\ref{fig:whitebox_log} highlights the comparison between a white-box and a black-box adversary. The results show that using gradient norms will improve the overall AUC both for our online attack, as well as when using a global threshold as in the LOSS attack.
However, at lower false-positive rates we do not observe any improvement of using gradient norms compared to just using model confidences. 

\section{Additional Figures and Tables}

\subsection{Attack Performance versus Model Accuracy}

\begin{figure}[h]
    \centering
    \includegraphics[scale=.6]{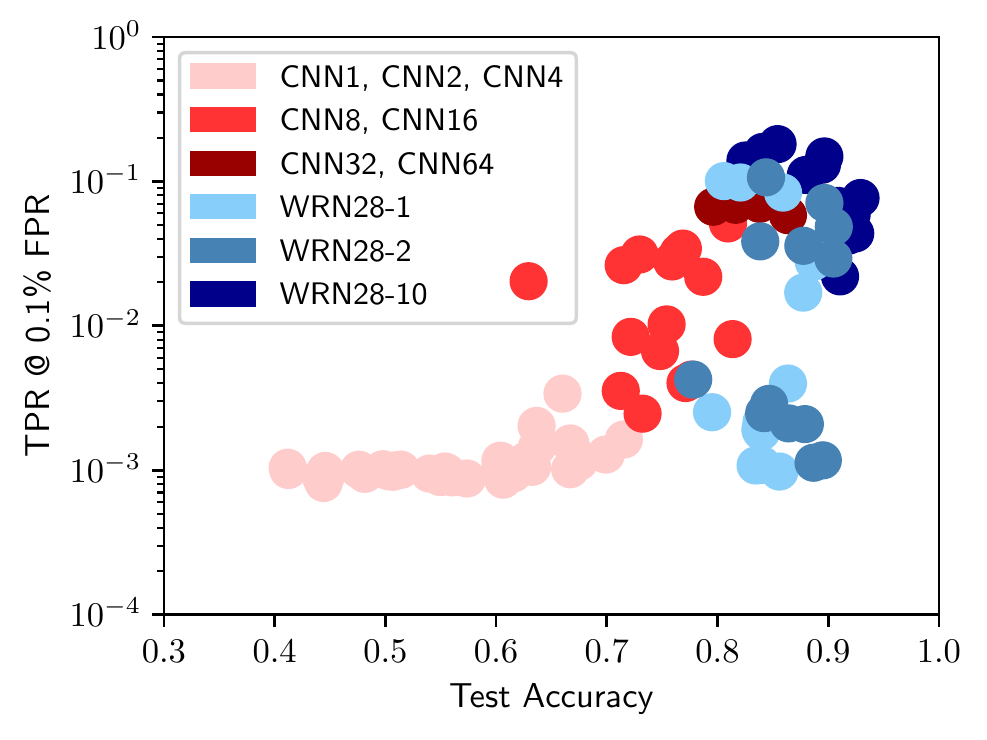}
    \caption{Attack true-positive rate versus model test accuracy.}
    \label{fig:tpr0p1_test}
\end{figure}

In \Cref{ssec:overfitting}, \Cref{fig:tpr0p1_gap} we plotted the relationship between a model's train-test gap and its vulnerability to membership inference attacks.
In \Cref{fig:tpr0p1_test}, we look at the attack success rate as a function of the \emph{test accuracy} of the same models.
There is a clear trend where \emph{better models are more vulnerable to attacks}. Prior work reported a similar phenomenon for data extraction attacks~\cite{carlini2019secret, carlini2020extracting}.

\subsection{Full ROC Curves for Gaussian Distribution Fitting}

\begin{figure}[h]
    \centering
    \includegraphics[scale=.6]{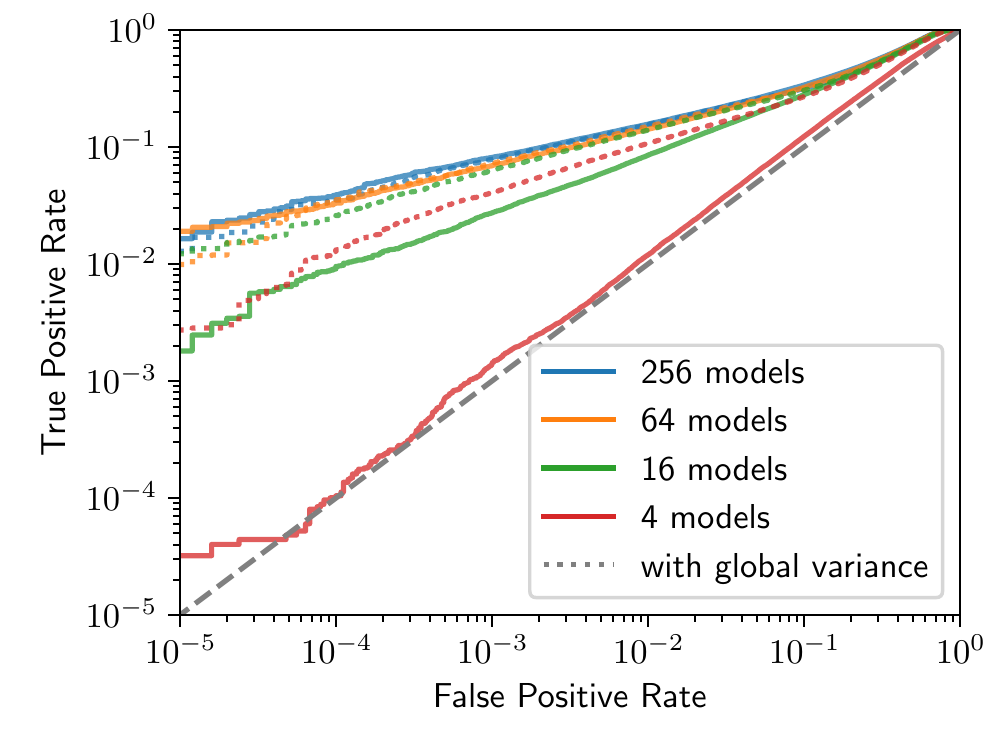}
    \caption{Effect of varying the number of models trained on attack success rates.
    It is always useful to estimate the mean per-example difficulty;
    however when only a few models are available,
    it is orders of magnitude more effective to assign all examples the same
    variance.
    }
    \label{fig:gaussian_fit_full}
\end{figure}

In \Cref{fig:gaussian_fit_full}, we show full (log-scale) ROC curves for the experiment in \Cref{ssec:global_variance}, where we explored the effect of varying the number of shadow models on the success rate of our online attack.
We vary the number of shadow models from $4$ to $256$ and consider two attack variants: (1) fit Gaussians for each example by estimating the means $\mu_{\text{in}}, \mu_{\text{out}}$ and variances $\sigma_{\text{in}}^2, \sigma_{\text{out}}^2$ independently for each example; (2) estimate the means $\mu_{\text{in}}, \mu_{\text{out}}$ for each example, but estimate global variances $\sigma_{\text{in}}^2, \sigma_{\text{out}}^2$. As we observed in \Cref{ssec:global_variance}, estimating per-example variances works poorly when the number of shadow models is small ($<64$). With a global estimate of the variance, the attack performs nearly on par with our best attack with as little as 16 shadow models. 

\clearpage

\subsection{Comparison to Prior Work on Additional Datasets}

\begin{figure}[h]
    \centering
    \centering
    \includegraphics[scale=0.6]{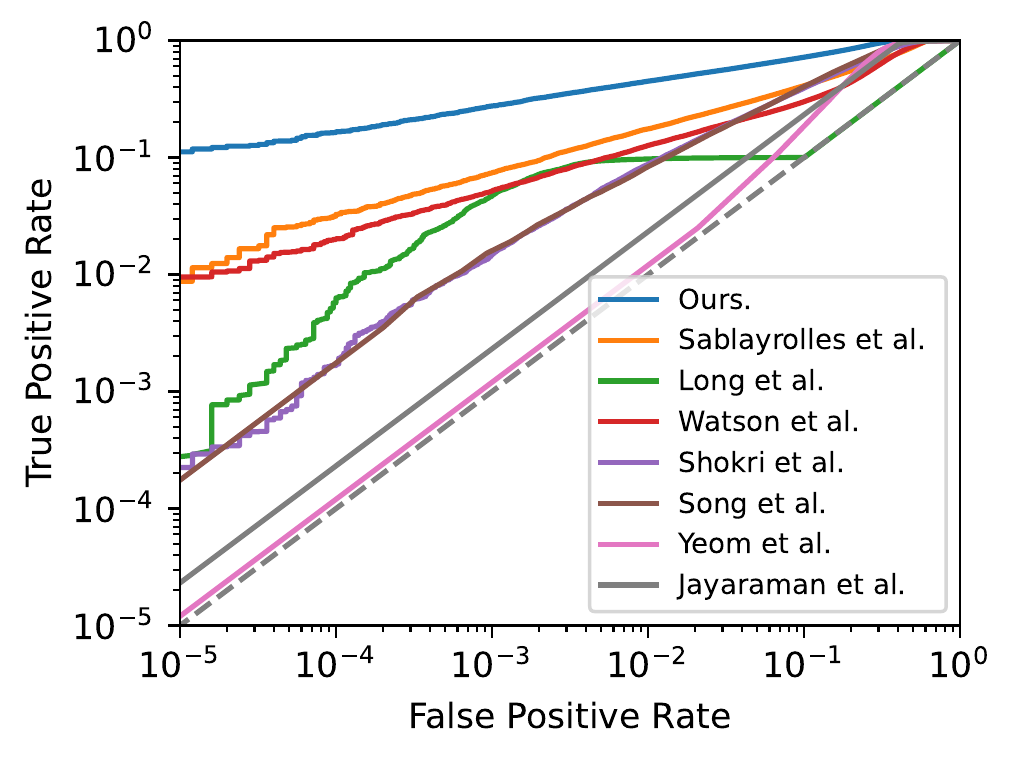}
    \caption{ROC curve of prior membership inference attacks, compared to our attack, on CIFAR-100.}
    \label{fig:cifar100_dataset}
\end{figure}

\begin{figure}[h]
    \centering
    \includegraphics[scale=0.6]{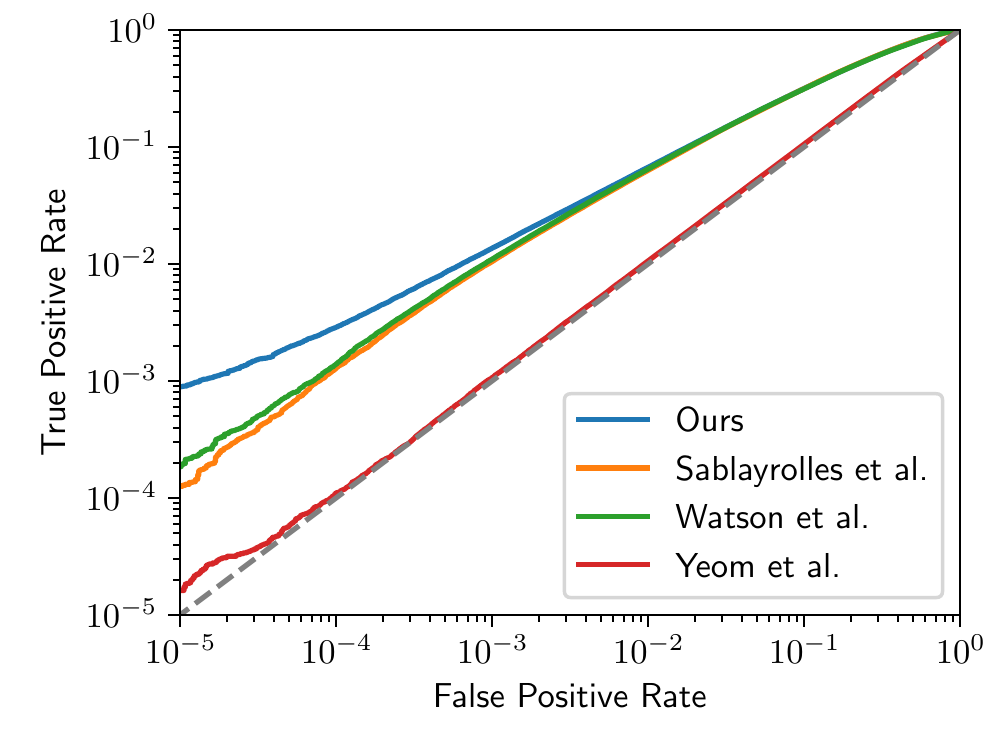}
    \caption{ROC curve of prior membership inference attacks, compared to our attack, on WikiText-103. We omit prior attacks that rely on the model features $z(x)$, as these attacks were not designed for sequential models.\\}
    \label{fig:wikitext_dataset}
\end{figure}

\begin{figure}[t]
    \centering
    \includegraphics[scale=0.6]{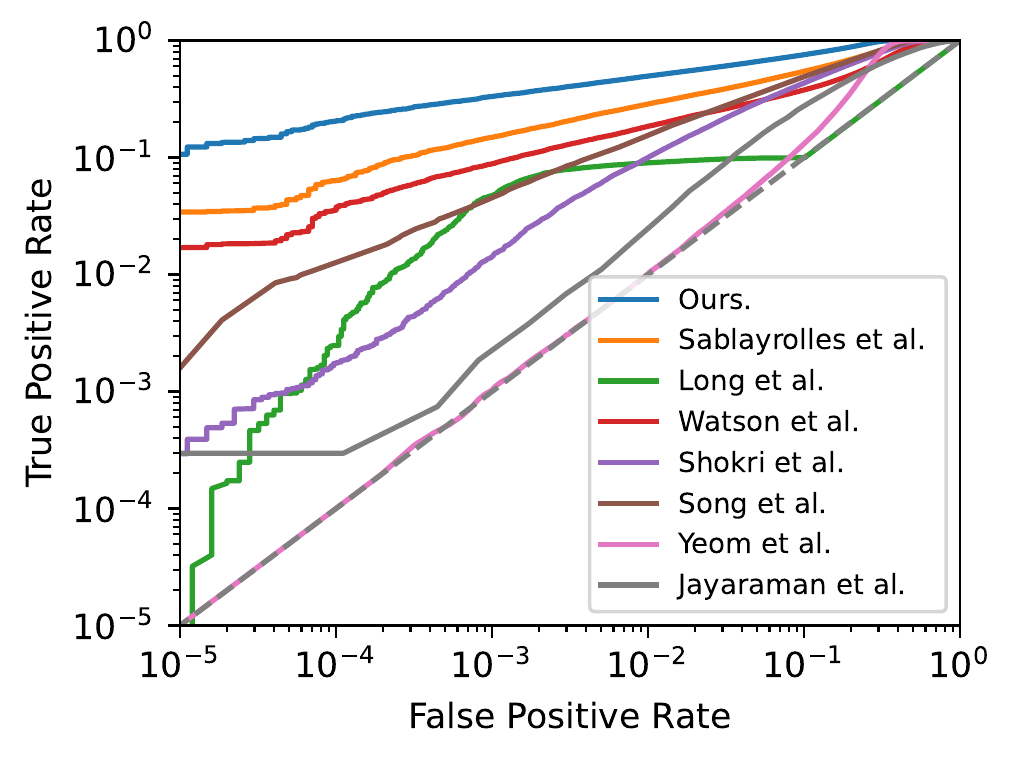}
    \caption{ROC curve of prior membership inference attacks, compared to our attack, on the Texas dataset.\\}
    \label{fig:texas_dataset}
%
    \centering
    \includegraphics[scale=0.6]{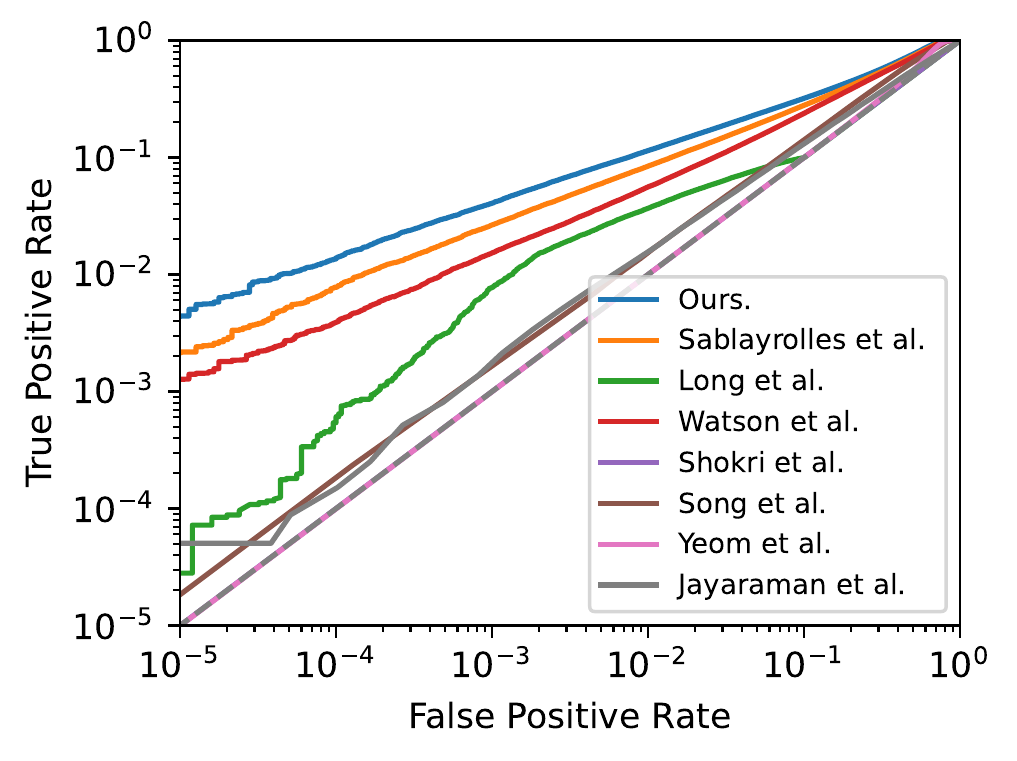}
     \caption{ROC curve of prior membership inference attacks, compared to our attack, on the Purchase dataset.}
    \label{fig:purchase_dataset}
\end{figure}

Similarly to \Cref{fig:full_comparison} for CIFAR-10, we compare our attack against prior membership inference attacks on additional datasets: CIFAR-100 in \Cref{fig:cifar100_dataset}, WikiText-103 in \Cref{fig:wikitext_dataset}, Texas in \Cref{fig:texas_dataset} and Purchase in \Cref{fig:purchase_dataset}.

\newpage
\vphantom{dummy text so that figures are nicely top-aligned}

\newpage
\vphantom{dummy text so that figures are nicely top-aligned}
\subsection{Attack Ablations on Additional Datasets}

Similarly to \Cref{tab:ablation_summary} for CIFAR-10, we now perform ablations on the different components of our attack for CIFAR-100 (\Cref{tab:ablation_cifar100}), WikiText-103 (\Cref{tab:ablation_wikitext}), Texas (\Cref{tab:ablation_texas}) and Purchase (\Cref{tab:ablation_purchase}).
Note that for WikiText-103, Texas and Purchase, we train models without any data augmentations and thus do not perform augmentations in the attack either.

As we observed in \Cref{sec:ablation} for CIFAR-10, a Gaussian Likelihood Test after logit scaling significantly boosts the performance of past attacks that rely on per-example thresholds, both in the offline case and in the online case.

In contrast to CIFAR-10, we observe that logit scaling on its own is often \emph{detrimental} to the online attack of \citet{sablayrolles2019white}. Similarly, we find that a Gaussian Likelihood Test on its own (i.e., without logit scaling) often hurts the attack performance. Thus, these two components necessarily have to be applied \emph{together} to achieve a good attack performance.

\begin{table}[t]
    \footnotesize
    \centering
    \begin{tabular}{@{}lr@{}}
    \toprule
        \textbf{Attack Approach} & \textbf{TPR @ 0.1\% FPR} \\
        \midrule
        LOSS attack~\cite{yeom2018privacy} & 0.0\% \\
        \quad + Logit scaling & 0.1\% \\
        \quad + Multiple queries & 0.1\% \\
        \midrule
        LOSS attack~\cite{yeom2018privacy} & 0.0\% \\
        \quad + Per-example thresholds ($\tilde{\mathbb{Q}}_{\text{out}}$ only)~\cite{watson2021importance} & 5.2\% \\
        \quad + Logit scaling & 14.7\% \\
        \quad + Gaussian Likelihood & 18.9\% \\
        \quad + Multiple queries \textbf{(our offline attack)} & 22.3\% \\
        \midrule
        LOSS attack~\cite{yeom2018privacy} & 0.0\% \\
        \quad + Per-example thresholds ($\mathbb{Q}_{\text{in}}$ \& $\mathbb{Q}_{\text{out}}$)~\cite{sablayrolles2019white} & 7.4\% \\
        \quad + Logit scaling & 2.8\% \\
        \quad + Gaussian Likelihood & 24.1\% \\
        \quad + Multiple queries \textbf{(our attack)} & 27.6\% \\
    \bottomrule
    \end{tabular}
    \caption{Breakdown of how various components build up to obtain our best attacks on the CIFAR-100 dataset.\\}
    \label{tab:ablation_cifar100}
\end{table}

\begin{table}[t]
    \footnotesize
    \centering
    \begin{tabular}{@{}lr@{}}
    \toprule
        \textbf{Attack Approach} & \textbf{TPR @ 0.1\% FPR} \\
        \midrule
        LOSS attack~\cite{yeom2018privacy} & 0.1\% \\
        \quad + Logit scaling & 0.1\% \\
        \midrule
        LOSS attack~\cite{yeom2018privacy} & 0.1\% \\
        \quad + Per-example thresholds ($\tilde{\mathbb{Q}}_{\text{out}}$ only)~\cite{watson2021importance} & 1.1\% \\
        \quad + Logit scaling & 1.1\% \\
        \quad + Gaussian Likelihood \textbf{(our offline attack)} & 1.2\% \\
        \midrule
        LOSS attack~\cite{yeom2018privacy} & 0.1\% \\
        \quad + Per-example thresholds ($\mathbb{Q}_{\text{in}}$ \& $\mathbb{Q}_{\text{out}}$)~\cite{sablayrolles2019white} & 1.0\% \\
        \quad + Logit scaling & 1.0\% \\
        \quad + Gaussian Likelihood \textbf{(our attack)} & 1.4\% \\
    \bottomrule
    \end{tabular}
    \caption{Breakdown of how various components build up to obtain our best attacks on the WikiText-103 dataset.\\}
    \label{tab:ablation_wikitext}
\end{table}


\subsection{Full ROC Curves for Mismatched Training Procedures}

\begin{figure*}[h]
\centering

\begin{subfigure}[b]{0.32\textwidth}
\centering
\includegraphics[width=\textwidth]{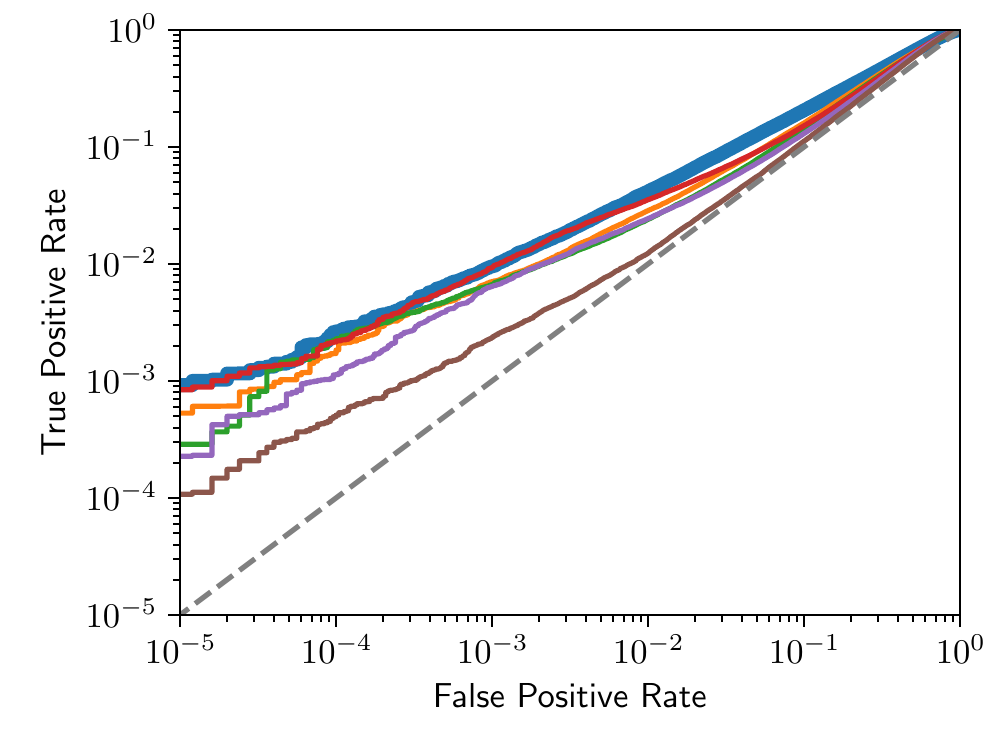}
\caption{CNN-16 as target.}
\end{subfigure}
\begin{subfigure}[b]{0.32\textwidth}
\centering
\includegraphics[width=\textwidth]{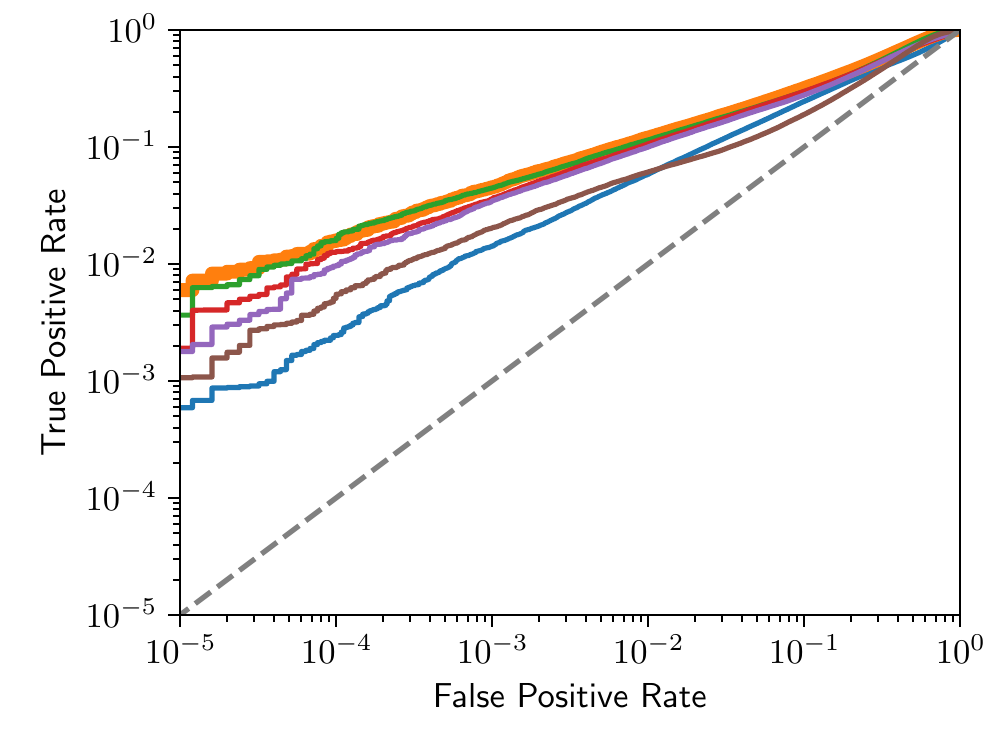}
\caption{CNN-32 as target.}
\end{subfigure}
\begin{subfigure}[b]{0.32\textwidth}
\centering
\includegraphics[width=\textwidth]{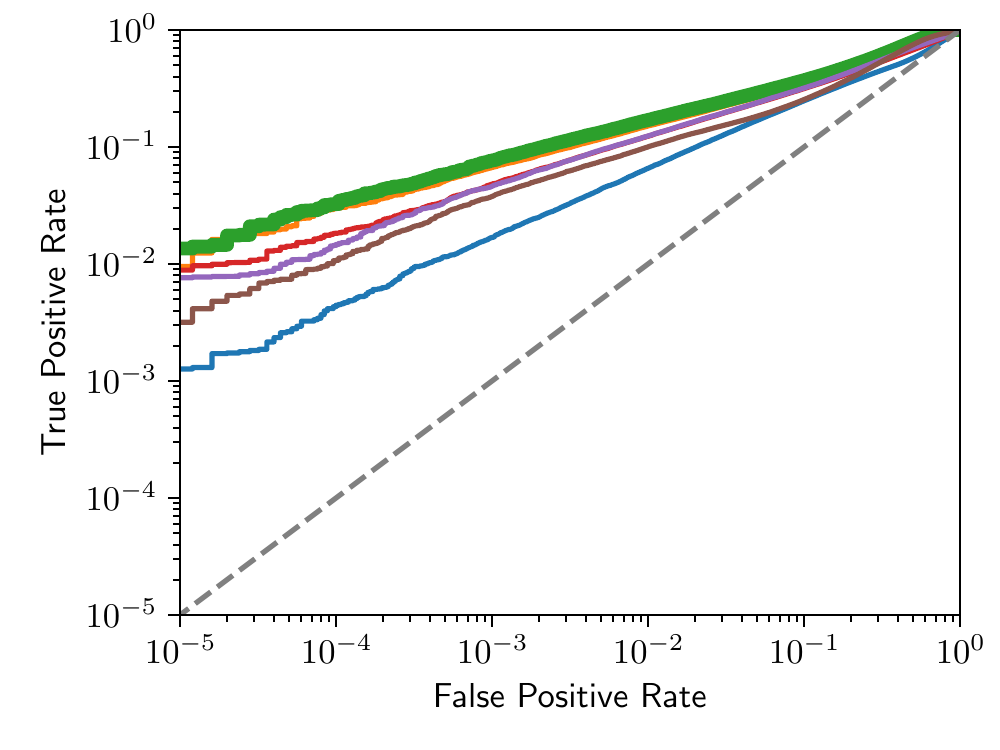}
\caption{CNN-64 as target.}
\end{subfigure}
\\
\begin{subfigure}[b]{0.32\textwidth}
\centering
\includegraphics[width=\textwidth]{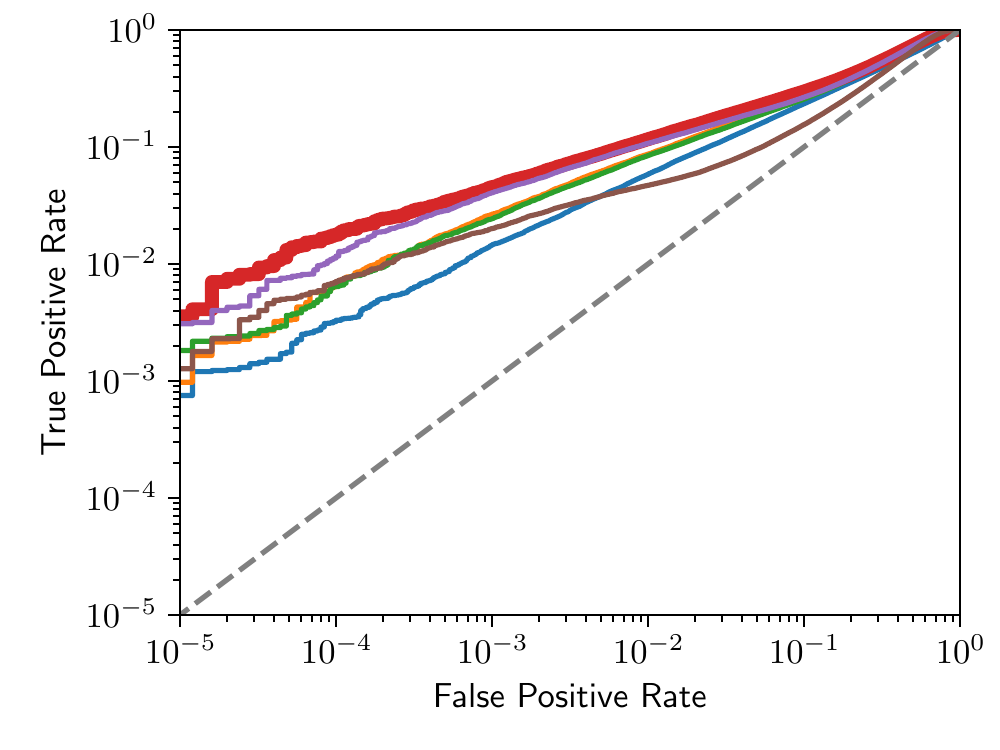}
\caption{WRN28-1 as target.}
\end{subfigure}
\begin{subfigure}[b]{0.32\textwidth}
\centering
\includegraphics[width=\textwidth]{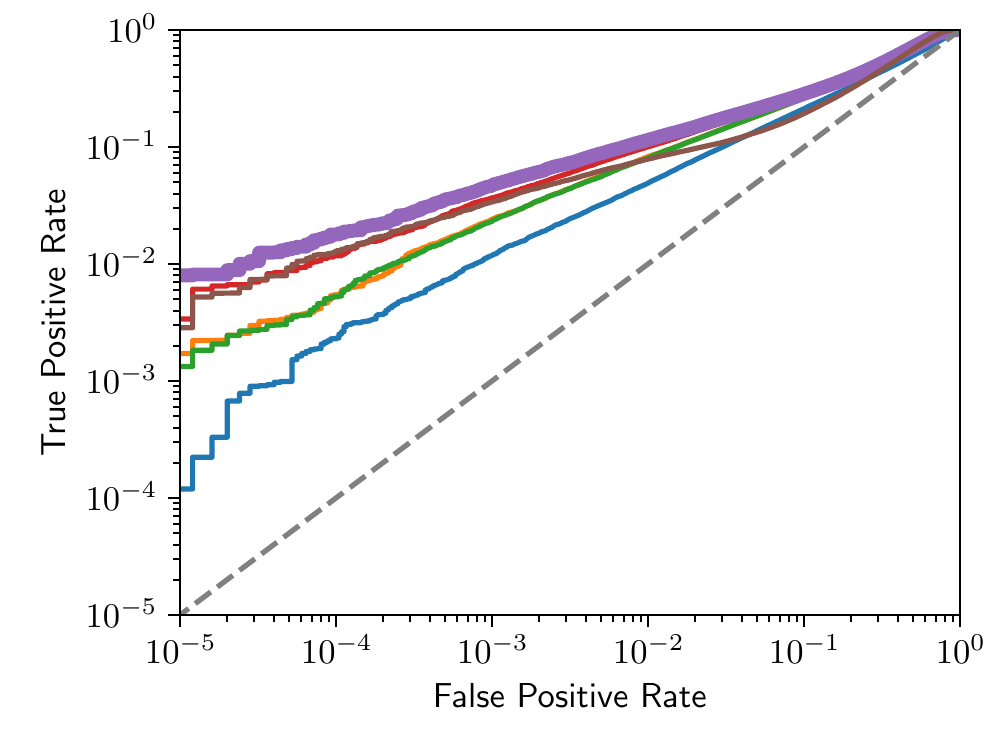}
\caption{WRN28-2 as target.}
\end{subfigure}
\begin{subfigure}[b]{0.32\textwidth}
\centering
\includegraphics[width=\textwidth]{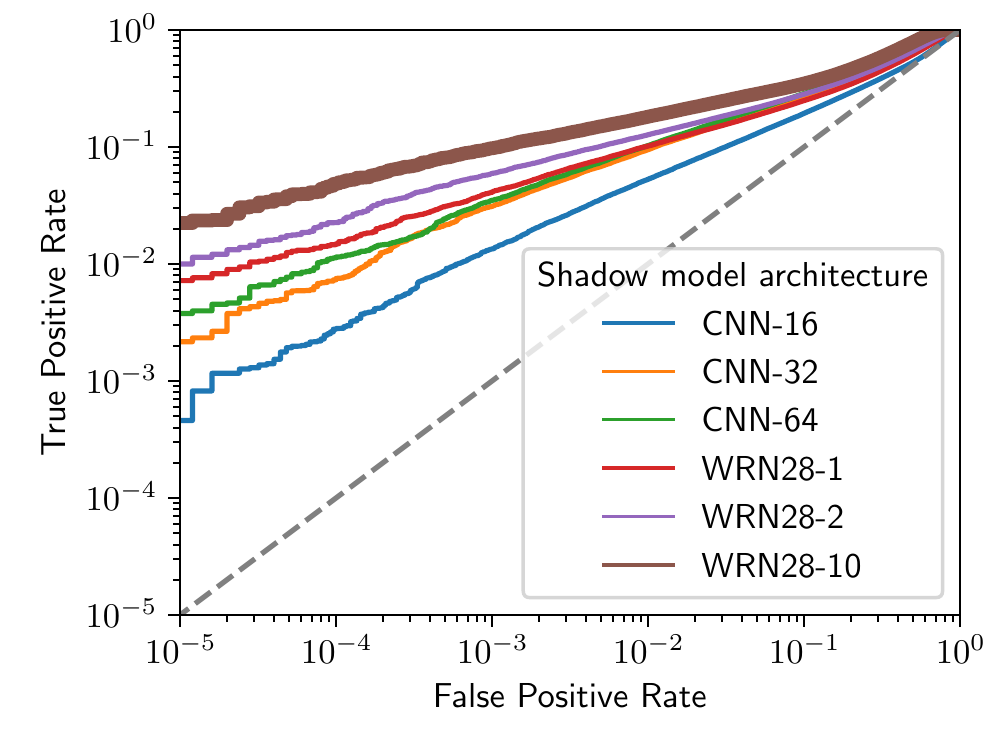}
\caption{WRN28-10 as target.}
\end{subfigure}

\caption{Different architectures with momentum optimizer and mirror \& shift as augmentation.}
\label{fig:architecture}
\end{figure*}

\begin{figure*}[h]
\centering
\begin{subfigure}[b]{0.32\textwidth}
\centering
\includegraphics[width=\textwidth]{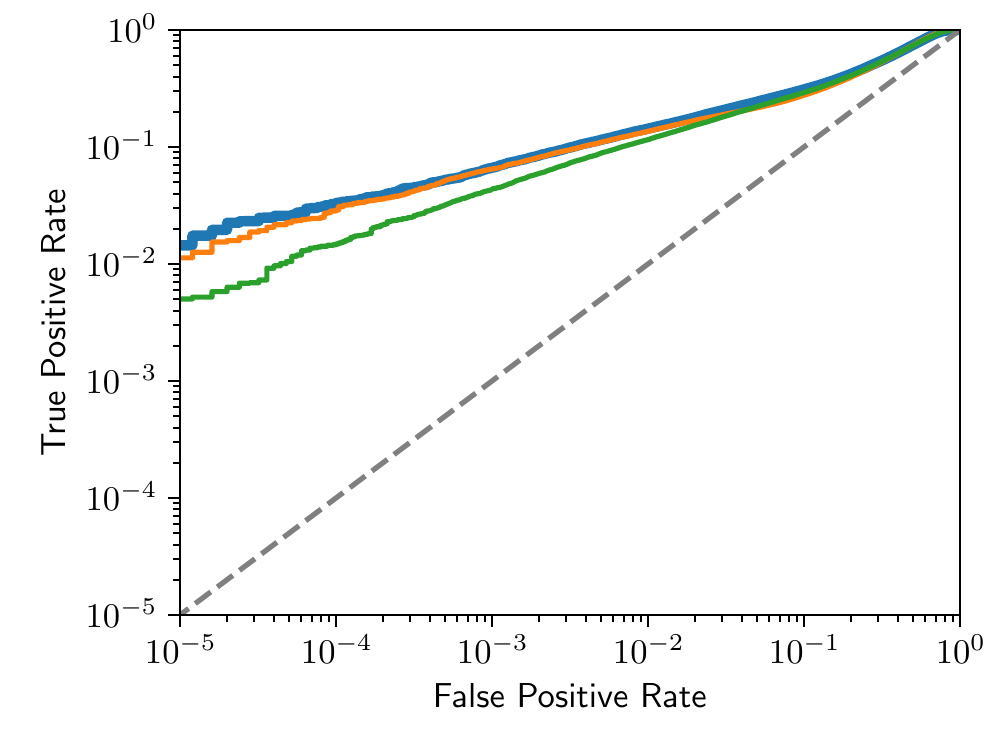}
\caption{SGD as target.}
\end{subfigure}
\begin{subfigure}[b]{0.32\textwidth}
\centering
\includegraphics[width=\textwidth]{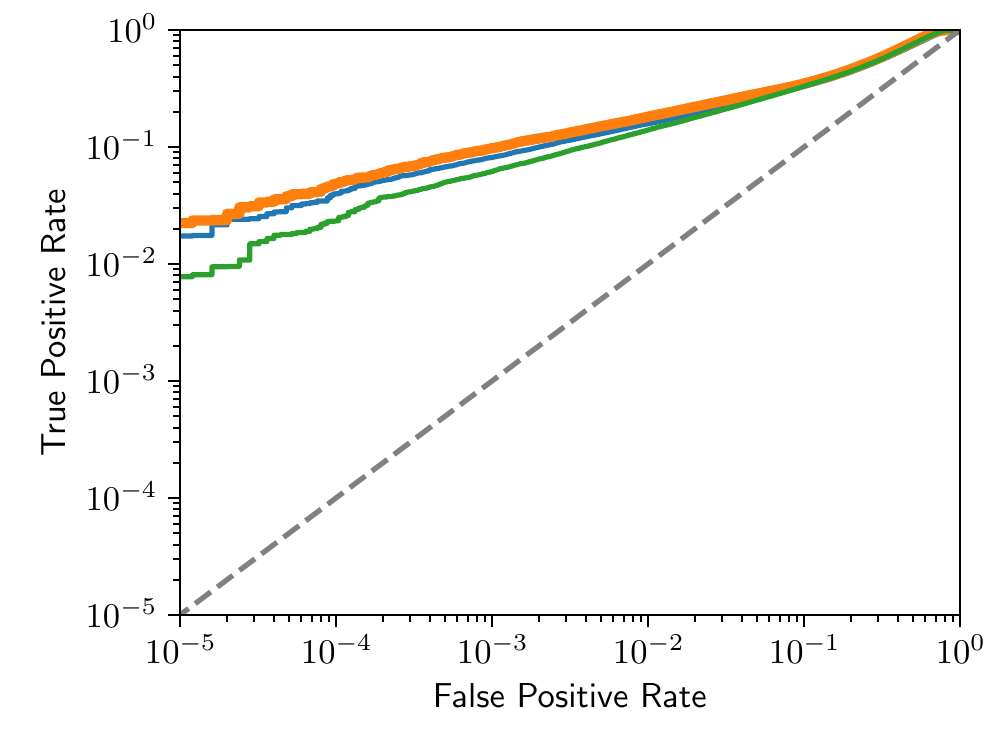}
\caption{Momentum as target.}
\end{subfigure}
\begin{subfigure}[b]{0.32\textwidth}
\centering
\includegraphics[width=\textwidth]{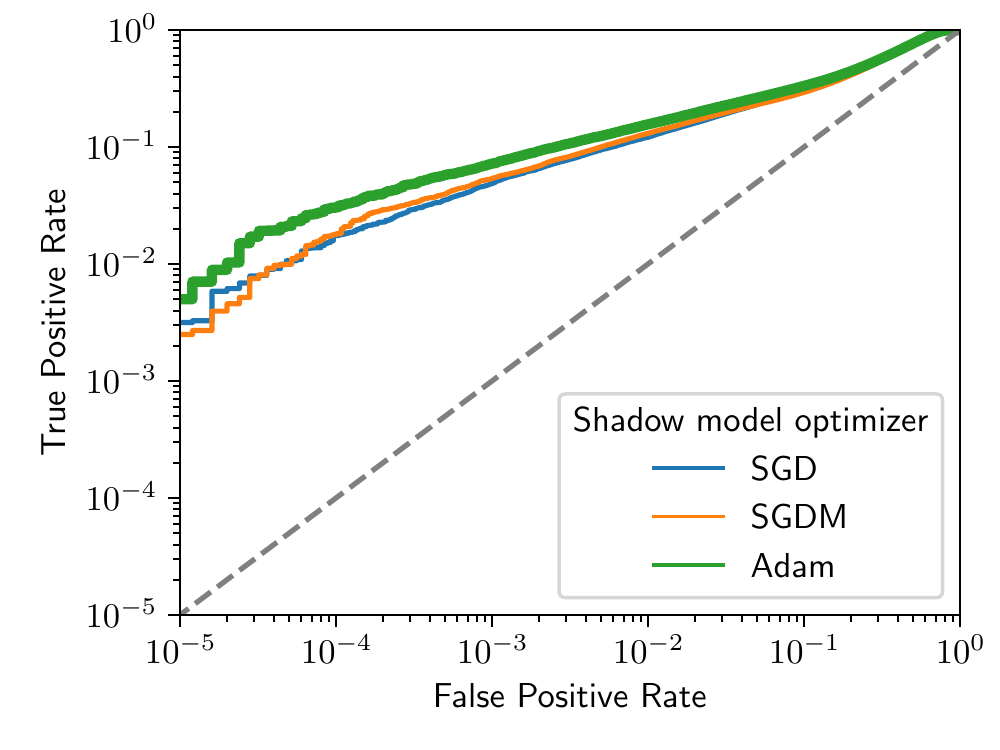}
\caption{Adam as target.}
\end{subfigure}
\caption{Different optimizers on WRN28-10 with mirror \& shift as augmentation.}
\label{fig:wrn28_optimizer}
\end{figure*}

\begin{figure*}[ht]
\centering

\begin{subfigure}[b]{0.24\textwidth}
\centering
\includegraphics[width=\textwidth]{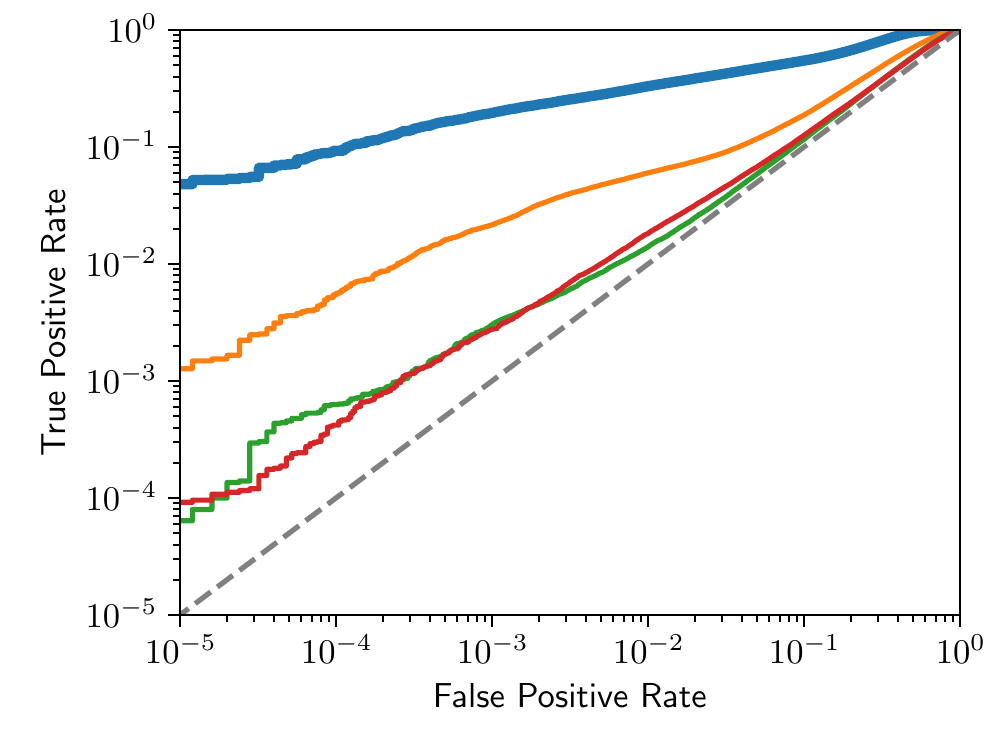}
\caption{No augmentation.}
\end{subfigure}
\begin{subfigure}[b]{0.24\textwidth}
\centering
\includegraphics[width=\textwidth]{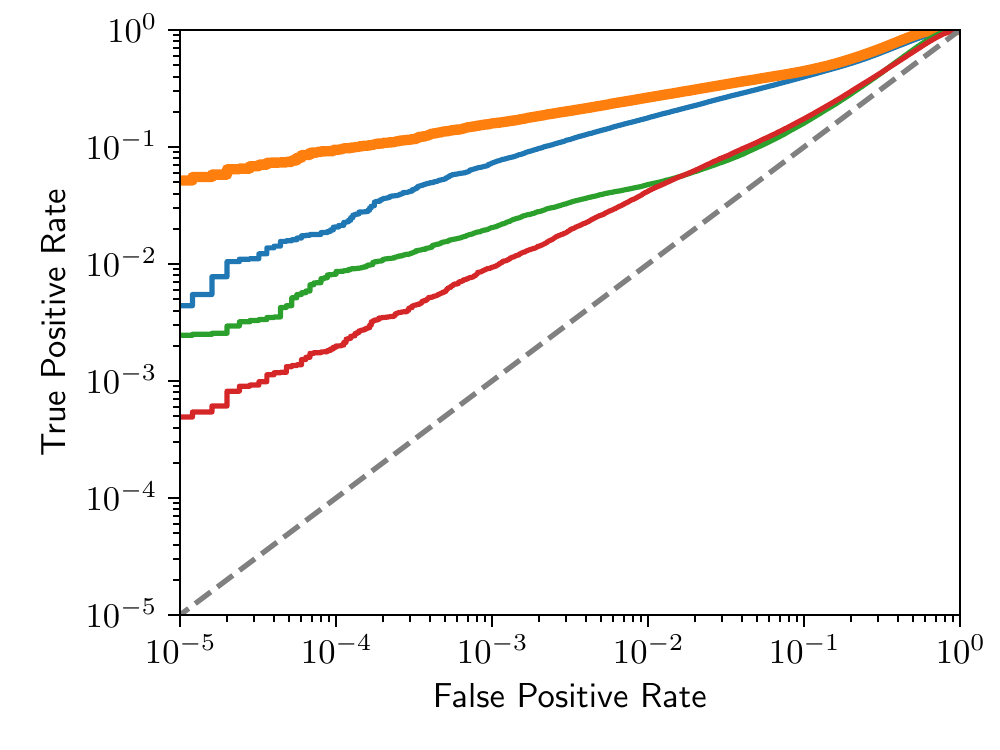}
\caption{Mirror.}
\end{subfigure}
\begin{subfigure}[b]{0.24\textwidth}
\centering
\includegraphics[width=\textwidth]{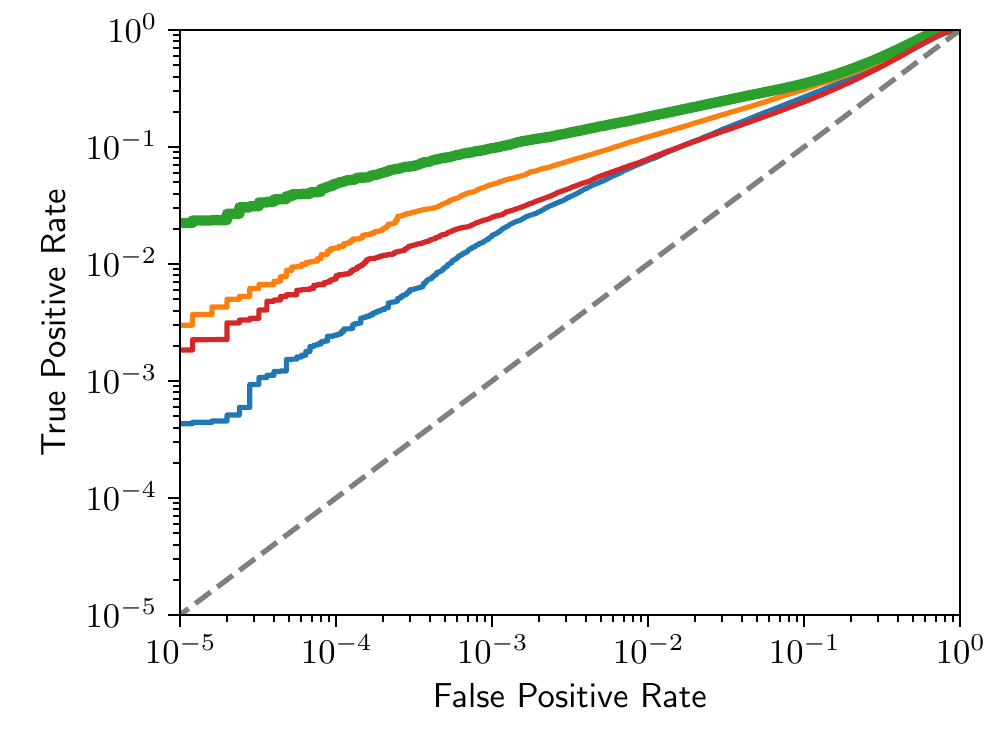}
\caption{Mirror+Shift.}
\end{subfigure}
\begin{subfigure}[b]{0.24\textwidth}
\centering
\includegraphics[width=\textwidth]{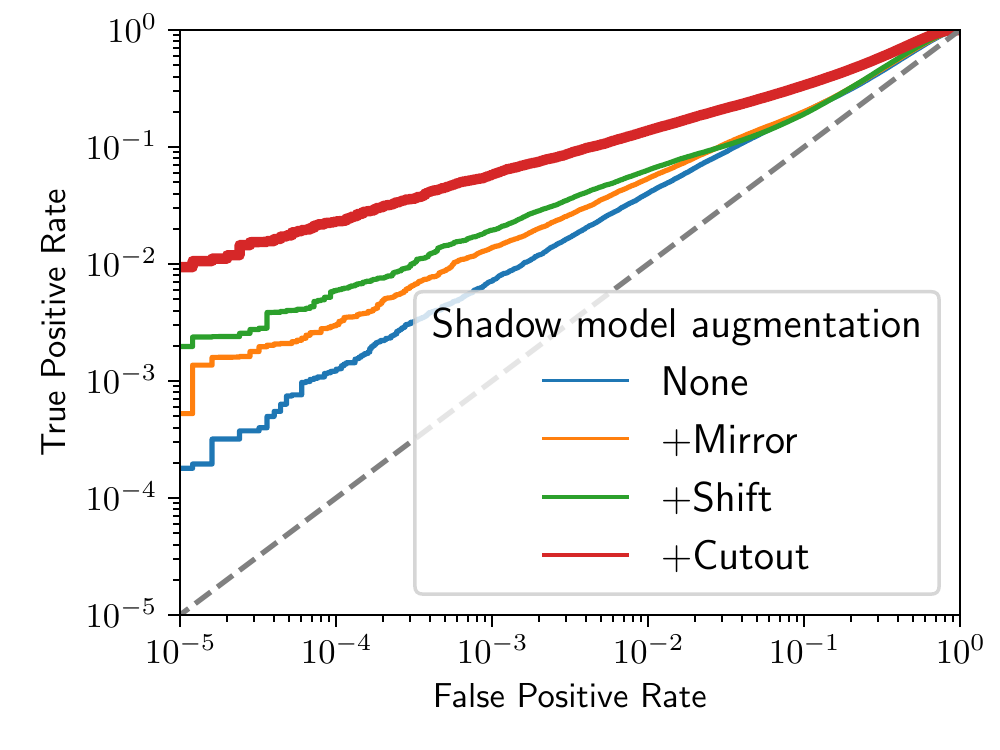}
\caption{Mirror+Shift+Cutout16.}
\end{subfigure}

\caption{Different augmentations on WRN28-10 with momentum optimizer.}
\label{fig:wrn28_augmentation}
\end{figure*}

In \Cref{fig:architecture,fig:wrn28_optimizer,fig:wrn28_augmentation}, we plot full ROC curves for the experiments from \Cref{ssec:mismatched_training}, where the attacker has to guess the architecture, optimizer and data augmentation used by the target model.

\begin{table}[t]
    \footnotesize
    \centering
    \begin{tabular}{@{}lr@{}}
    \toprule
        \textbf{Attack Approach} & \textbf{TPR @ 0.1\% FPR} \\
        \midrule
        LOSS attack~\cite{yeom2018privacy} & 0.1\% \\
        \quad + Logit scaling & 0.1\% \\
        \midrule
        LOSS attack~\cite{yeom2018privacy} & 0.1\% \\
        \quad + Per-example thresholds ($\tilde{\mathbb{Q}}_{\text{out}}$ only)~\cite{watson2021importance} & 8.8\% \\
        \quad + Logit scaling & 19.0\% \\
        \quad + Gaussian Likelihood \textbf{(our offline attack)} & 24.6\% \\
        \midrule
        LOSS attack~\cite{yeom2018privacy} & 0.1\% \\
        \quad + Per-example thresholds ($\mathbb{Q}_{\text{in}}$ \& $\mathbb{Q}_{\text{out}}$)~\cite{sablayrolles2019white} & 14.9\% \\
        \quad + Logit scaling & 8.4\% \\
        \quad + Gaussian Likelihood \textbf{(our attack)} & 33.2\% \\
    \bottomrule
    \end{tabular}
    \caption{Breakdown of how various components build up to obtain our best attacks on the Texas dataset.\\}
    \label{tab:ablation_texas}
\end{table}

\begin{table}[t]
    \footnotesize
    \centering
    \begin{tabular}{@{}lr@{}}
    \toprule
        \textbf{Attack Approach} & \textbf{TPR @ 0.1\% FPR} \\
        \midrule
        LOSS attack~\cite{yeom2018privacy} & 0.0\% \\
        \quad + Logit scaling & 0.1\% \\
        \midrule
        LOSS attack~\cite{yeom2018privacy} & 0.0\% \\
        \quad + Per-example thresholds ($\tilde{\mathbb{Q}}_{\text{out}}$ only)~\cite{watson2021importance}  & 1.5\% \\
        \quad + Logit scaling & 1.3\% \\
        \quad + Gaussian Likelihood \textbf{(our offline attack)} & 1.4\% \\
        \midrule
        LOSS attack~\cite{yeom2018privacy} & 0.0\% \\
        \quad + Per-example thresholds ($\mathbb{Q}_{\text{in}}$ \& $\mathbb{Q}_{\text{out}}$)~\cite{sablayrolles2019white} & 2.7\% \\
        \quad + Logit scaling & 0.2\% \\
        \quad + Gaussian Likelihood \textbf{(our attack)} & 4.1\% \\
    \bottomrule
    \end{tabular}
    \caption{Breakdown of how various components build up to obtain our best attacks on the Purchase dataset.}
    \label{tab:ablation_purchase}
\end{table}

\vphantom{dummy text so that tables are nicely top-aligned}
\clearpage

\end{document}